\documentclass{personal}

\usepackage{amssymb}
\usepackage{amsmath}

\newcommand{\bra}[1]{\langle #1|}
\newcommand{\ket}[1]{|#1\rangle}
\newcommand{\braket}[2]{\langle #1|#2\rangle}

\def\Nj{\{N_j\}}

\def\d{{\rm d}}
\def\i{i}
\def\parti{n} %indice delle particelle%
\def\Pro{{\sf P}}

\newcommand{\tr}{{\rm tr}}

\newcommand{\iotav}{\boldsymbol{\iota}}
\newcommand{\rhov}{\boldsymbol{\rho}}
\newcommand{\etav}{\boldsymbol{\eta}}
\newcommand{\phiv}{\boldsymbol{\phi}}

\newcommand{\Psis}{{\sf \Psi}}

\newcommand{\poinc}{IO(1,3)$^\uparrow$}

\begin{document}

\begin{frontmatter}



%********************************************************************************
\title{The microcanonical ensemble of the ideal relativistic quantum gas
with angular momentum conservation}
%********************************************************************************


\author{F. Becattini, L. Ferroni}

\address{Universit\`a di Firenze and INFN Sezione di Firenze}

\begin{abstract}
We derive the microcanonical partition function of the ideal relativistic 
quantum gas with fixed intrinsic angular momentum as an expansion over fixed 
multiplicities. We developed a group theoretical approach by generalizing 
known projection techniques to the Poincar\'e group. Our calculation is 
carried out in a quantum field framework and applies to particles with any 
spin. It extends known results in literature in that it does not introduce 
any large volume approximation and it takes particle spin fully into account. 
We provide expressions of the microcanonical partition function at fixed 
multiplicities in the limiting classical case of large volumes and large 
angular momenta and in the grand-canonical ensemble. We also derive the 
microcanonical partition function of the ideal relativistic quantum gas with 
fixed parity. 
\end{abstract}

\end{frontmatter}

%**************************************************************************
\section{Introduction}
%**************************************************************************

In a previous work \cite{micro1}, we derived the microcanonical partition function 
of an ideal relativistic quantum gas of spinless bosons within a quantum field
theoretical framework. Therein, we confined ourselves to the microcanonical 
ensemble with fixed energy and momentum. Here, we will obtain the partition 
function of the microcanonical ensemble of an ideal relativistic quantum gas with 
fixed energy-momentum {\em and} angular momentum. It should be pointed out from the
very beginning that, for a quantum system, energy-momentum and angular momentum can
be defined simultaneously only in the rest frame, i.e. if ${\bf P}=0$. In fact, more
rigorously, one should rather speak of fixing the maximal 
set of commuting observables of the orthochronous Poincar\'e group \poinc, including
energy-momentum and two observables formed out of the Pauli-Lubanski vector. However, 
for the sake of simplicity, we will retain the more colloquial expressions of 
``energy-momentum and angular momentum conservation'' or ``microcanonical ensemble
with angular momentum conservation" being understood that we are working in the system
rest frame, where the Pauli-Lubanski vector is proportional to the intrinsic angular 
momentum or spin of the system.

The microcanonical partition function of an ideal relativistic quantum gas with 
fixed angular momentum, in a full quantum field theoretical framework, has never been 
derived for particles with spin. It has been obtained in an essentially non-relativistic
quantum mechanical framework in ref.~\cite{cerang} and an equivalent calculation
has been performed in ref.~\cite{satz} by using a projection method, yet only for 
spinless particles. With the same limitations, an expression with angular momentum 
conservation, but without momentum conservation has been derived by Koba~\cite{koba}. 
The classical limit has been worked out in ref.~\cite{ericson}.
In this work we will get over these limitations and derive the microcanonical 
partition function of the ideal relativistic quantum gas of particles with spin by
using a group theoretical approach including the maximal set of observables 
pertaining to space-time symmetries, i.e. energy-momentum, spin, and parity, in 
a consistent way in a quantum field framework. Like in ref.~\cite{micro1}, we will 
write it as an expansion over fixed multiplicities. 

The most general expression will be used to work out special cases, such as the 
grand-canonical (fixed temperature) and the classical limit. We will show that those, 
more familiar, expressions, can be recovered, thus supporting the correctness of our 
approach.   

%==================================================================
\subsection*{Notation}
%==================================================================

In this paper we adopt the natural units, with $\hslash=c=K=1$. Space-time 
linear transformations (translations, rotations, boosts) and SL(2,C) transformations
are written in serif font, e.g. ${\sf R}$, ${\sf L}$. Operators in Hilbert space will
be denoted by a hat, e.g. $\hat {\sf R}$. The only exception to this rule are field
operators, which will be written as a capital $\Psi$, while c-numbers field functions 
with a small $\psi$. Also unit vectors will be denoted with an upper hat, but 
they will always be referred to explicitely in the text.

%****************************************************************************
\section{The projection onto irreducible Poincar\'e states}
%****************************************************************************

In order to calculate the microcanonical partition function of an ideal relativistic 
quantum gas fixing the maximal set of observables pertaining to space-time symmetries, 
the use of group theory is both essential and enlightening. The statement of the problem 
in this framework has been developed in refs.~\cite{bf1,meaning}, generalizing known 
projection operator techniques used for internal, rather than space-time, symmetries,
and we will outline it here. 
 
In ref.~\cite{micro1} we showed that the microcanonical partition function of a 
relativistic system can be written as:
\begin{equation}\label{micropf1}
 \Omega =  \e^S = \sum_{h_V} \bra{h_V} \Pro_i \ket{h_V} \; , 
\end{equation}
where $\ket{h_V}$ are a complete set of states for the localized problem (the quantum
field in a compact region) and $\Pro_i$ is a projector operator which selects the 
quantum states with the fixed values of the physical observables; $S$ is, by definition,
the entropy. If only energy-momentum is involved, the relevant symmetry group is T(4) 
and the projector reads: 
\begin{equation}\label{t4}
 \Pro_i = \delta^4(P - \hat P) 
\end{equation}
Although $\delta^4(P -\hat{P})$ is not a proper projector because 
${\sf P}^2 = a{\sf P}$ where $a$ is a divergent constant\footnote{This is owing
to the fact that normalized projectors onto irreducible representations cannot be 
defined for non-compact groups, such as the space-time translation group T(4)} 
we will maintain this naming even for these non-idempotent operators relaxing
mathematical rigour.

Similarly, for the maximal set of observables, the projection $\Pro_i$ should be made 
onto irreducible states $i$ of the full symmetry group, namely the orthocronous 
Poincar\'e group \poinc. These states can be built by diagonalizing operators or 
combinations of operators of the Poincar\'e algebra. The first Casimir is the
squared four-momentum ${\hat P}^2$ with eigenvalue $M^2$. The construction of 
physical states depends on whether $M^2>0$ or $M^2=0$. We will consider only the 
former case, i.e. massive particles, and stick to the convention of 
ref.~\cite{moussa} for the definition of spin, that we will briefly outline.
 
First, the Pauli-Lubanski vector is formed:
\begin{equation}\label{luba}
\hat W_\mu = -(1/2) \sum_{\nu \rho \sigma} 
 \epsilon_{\mu \nu \rho \sigma} \hat J^{\nu \rho}  \hat P^\sigma 
\end{equation} 
$\hat J$ being the generators of the Lorentz group, with the following properties:
\begin{eqnarray}
&& [\hat W_\mu, \hat P_\nu] = 0 \\ \nonumber
&& [\hat W_\mu, \hat W_\nu] = -i \sum_{\rho \sigma} 
 \epsilon_{\mu \nu \rho \sigma} \hat W^\rho \hat P^\sigma \\ \nonumber
&& \hat W \cdot \hat P = 0
\end{eqnarray}
In view of the first commutation relation, if $\ket{p}$ is an eigenvector of 
$\hat P$, so is $\hat W \ket{p}$. The restriction of $\hat W$ to the
eigenspace labelled by four-momentum $P$ is defined as $\hat W(P)$. Since
$\hat W(P) \cdot P = 0$, this four-vector operator can be decomposed onto three 
orthonormal spacelike four-vectors $n_1(P),n_2(P),n_3(P)$ which form a basis of the 
Minkowski space with the unit vector $\hat P = P /\sqrt(P^2)$:
\begin{equation}\label{luba2}
  \hat W(P) = \sum_{i=1}^3 \hat W_i(P) n_i(P)
\end{equation}
It can be shown that the operators:
\begin{equation}\label{spins}
  \hat S_i(P) = \hat W_i(P)/M
\end{equation}
form an SU(2) algebra and are the suitable relativistic generalization of the spin
angular momentum. The third component $\hat S_3(P)$ can be diagonalized along with  
$\hat S^2 = -\hat W^2/M^2$ which is a Casimir of the full group \poinc,
with corresponding eigenvalues $\lambda$ and $J(J+1)$. With a suitable choice of
$n_i(P)$, i.e.:
\begin{equation}\label{choi}
  n_i (P) = [P]e_i \qquad [P] \equiv {\sf R}_3(\varphi) {\sf R}_2 (\theta) 
 {\sf L}_3(\xi)
\end{equation}
$e_i$ being the unit vectors of spacial axes and $[P]$ being a Lorentz 
transformation bringing the timelike vector $P_0 = (M,0,0,0)$ into the 
four-momentum $P$ with polar coordinates $\xi,\theta,\varphi$; the eigenvalue $\lambda$
has the physical meaning of the component of intrinsic angular momentum in the
rest frame along the direction of particle momentum ${\bf P}$. Thus, with the
choice (\ref{choi}), $\lambda$ is the helicity in the rest frame and $J$ is, by
definition, {\em the} spin of the particle. Since, from eqs.~(\ref{spins}), 
(\ref{luba2}) and (\ref{luba}) $\hat S_i(P_0) = \hat J_i$, the spin
operators in the rest frame coincide with the generators of the rotation groups.
    
The inclusion of parity adds another discrete quantum number $\Pi$ to the maximal set
of eigenvalues. Altogether, an irreducible state of the this group can be labelled 
by a four-momentum $P$, a spin $J$, its third component $\lambda$ and parity $\Pi$ 
and the projector onto such state $\Pro_i$ can then be written as:
\begin{equation}\label{progen}
   {\sf P}_i = {\sf P}_{P J \lambda \Pi} 
\end{equation}

Generalizing the known integral expression for compact groups, one can formally
write the projector (\ref{progen}) as an integral over the Poincar\'e group:
\begin{equation}\label{proj}
 \Pro_{P J \lambda \Pi} = \frac{1}{2} \sum_{z={\sf I,\Pi}} \dim \nu 
 \int \d \mu(g_z) \; U^{\nu \dag}_{i  i}(g_z) \, \hat{g}_z   
\end{equation} 
where $\mu$ is the invariant group measure, $z$ is the identity or space inversion 
${\sf \Pi}$, $g_z \in$ \poinc, $U^\nu(g_z)$ is the matrix of the unitary irreducible 
representation $\nu$ the state $i$ belongs to and $\hat{g}_z$ is the unitary 
representation of $g_z$ in the Hilbert space. Like for the energy-momentum projector
(\ref{t4}), the expression (\ref{proj}) is not, strictly speaking, a projector because 
the group \poinc is non-compact and $\Pro_{P J \lambda \Pi}^2 = a \Pro_{P J \lambda \Pi}$ 
with $a$ divergent constant. However, we will keep calling (\ref{proj}) a projector 
relaxing mathematical rigor. Working in the rest frame of the system, 
with $P = (M, {\bf 0})$, the matrix element $U^{\nu\dag}_{i  i}(g_z)$ vanishes unless  
Lorentz transformations are pure rotations and this implies the reduction of the 
integration in (\ref{proj}) from IO(1,3)$^\uparrow$ to ${\rm T}(4)\otimes {\rm SU}(2) 
\otimes {\rm Z}_2$ \cite{bf1} (replacing SO(3) with its universal covering group SU(2)). 
The general transformation of the orthochronous Poincar\'e group $g_z$ may be then 
factorized as:
\begin{equation}
  g_z = {\sf T}(x) {\sf Z} {\sf \Lambda} = {\sf T}(x) {\sf Z} 
  {\sf L}_{\hat{\bf n}}(\xi) {\sf R}
\end{equation}
where ${\sf T}(x)$ is a translation by the four-vector $x$, ${\sf Z}$
is either the identity or the space inversion and ${\sf \Lambda} = 
{\sf L}_{\hat{\bf n}}(\xi) {\sf R}$ is a general orthochronous Lorentz transformation 
written as the product of a boost of hyperbolic angle $\xi$ along the space-like 
axis $\hat{\bf n}$ and a rotation $\sf R$ depending on three Euler angles. Thus
eq.~(\ref{proj}) becomes:
\begin{eqnarray}\label{propro}
 {\sf P}_{P J \lambda \Pi} &=& \frac{1}{2} \sum_{{\sf Z}={\sf I, \Pi}} 
 \frac{\dim \nu}{(2\pi)^4} \int \d^4 x \int \d {\sf \Lambda} \; U^\nu_{i  i}({\sf T}(x)
 {\sf Z}{\sf \Lambda})^{*} \, \hat{{\sf T}}(x) \hat{{\sf Z}} \; 
  \hat{{\sf \Lambda}} \nonumber \\
 &=& \frac{1}{2} \sum_{{\sf Z}={\sf I, \Pi}} 
  \frac{\dim \nu}{(2\pi)^4} \int \d^4 x \int \d {\sf \Lambda} \; 
 \e^{\i P \cdot x} \Pi^{z} U^\nu_{i  i}({\sf \Lambda})^{*} \; \hat{{\sf T}}(x) 
 \hat{{\sf Z}} \; \hat{{\sf \Lambda}}
\end{eqnarray}
where $z=0$ if ${\sf Z}= {\sf I}$ and $z=1$ if ${\sf Z}= {\sf \Pi}$.
In the above equation, the invariant measure $\d^4 x$ of the translation subgroup 
has been normalized with a coefficient $1/(2\pi)^4$ in order to yield a Dirac delta, 
as shown later. Furthermore, $\d {\sf \Lambda}$ is the invariant normalized 
measure of the Lorentz group, which can be written as \cite{lgroup}:
\begin{equation}\label{measure}
  \d {\sf \Lambda} = \d {\sf L}_{\bf n}(\xi) \, \d {\sf R} =
  \sinh^2 \xi \d \xi \, \frac{\d \Omega_{\hat{\bf n}}}{4 \pi} \, \d {\sf R}
\end{equation}
$\d {\sf R}$ being the invariant measure of SU(2) group normalized to 1, 
$\xi \in [0,+\infty)$ and $\Omega_{\hat{\bf n}}$ are the angular 
coordinates of the unit vector $\hat{\bf n}$. 

In the rest frame of the system, where $P=P_0=(M,{\bf 0})$, Lorentz transformations 
$\sf \Lambda$ comprising any non-trivial boost (i.e. with $\xi \ne 0$) would make 
the matrix element $U^\nu_{i  i}({\sf \Lambda})^{*}$ vanish. Therefore $\sf \Lambda$ 
reduces to the rotation ${\sf R}$ and (\ref{propro}) to:
\begin{equation}
 {\sf P}_{P J \lambda \Pi} = \frac{1}{2} \sum_{{\sf Z}={\sf I, \Pi}} \frac{1}{(2\pi)^4} 
 \int \d^4 x \; (2J + 1) \int \d {\sf R} \; \e^{\i P \cdot x} \Pi^{z}  
 \; D^{J}_{\lambda \lambda}({\sf R})^* \, \hat{{\sf T}}(x)\, \hat{{\sf Z}} 
 \,\hat{{\sf R}} \;
\end{equation} 
where we now use the ordinary symbol $D$ to denote matrices of SU(2) unitary 
representations. The irreducible representations of this group are now labelled 
by eigenvalues of its generators $\hat J^2$ and $\hat J_3$ which coincide
with the spin $J$ and its third component $\lambda$ defined above by means of operators
$\hat S_i(P)$. 
   
Taking into account that $[{\sf Z},{\sf R}] = 0$, we can move the $\hat{{\sf Z}}$ 
operator to the right of $\hat{{\sf R}}$ and recast the above equation as:
\begin{eqnarray}\label{final}
 {\sf P}_{P J \lambda \Pi} &=& \frac{1}{(2\pi)^4} 
 \int \d^4 x \; \e^{\i P \cdot x} \hat{{\sf T}}(x) (2J + 1) \int \d{\sf R} \; 
 D^{J}_{\lambda  \lambda}({\sf R}^{-1}) \,\hat{{\sf R}}  \, 
 \frac{{\sf I} + \Pi \hat{{\sf \Pi}}}{2}  \nonumber \\
   &=& \delta^4( P - \hat{P}) (2J + 1) \int \d{\sf R} \; 
 D^{J}_{\lambda  \lambda}({\sf R}^{-1}) \, \hat{{\sf R}} \, 
 \frac{{\sf I} + \Pi \hat{{\sf \Pi}}}{2}
\end{eqnarray} 
$\Pi$ being the parity of the system. 
The Eq.~(\ref{final}) is indeed the final general expression of the projector defining 
the proper microcanonical ensemble with $P=(M,{\bf 0})$, where all conservation laws 
related to space-time symmetries are implemented. The nice feature of the above 
expression is the factorization of projections onto the energy-momentum $P$, 
spin-helicity $J,\lambda$ and parity $\Pi$ which allows us to calculate the 
contribution of angular momentum and energy-momentum conservation separately. 

%*****************************************************************************
\section{The full microcanonical partition function}
%*****************************************************************************

The microcanonical partition function (MPF) (\ref{micropf1}) can be written also 
as, by inserting a resolution of identity $\sum_f \ket{f}\bra{f}$: 
\begin{equation}\label{micropf2}
 \Omega = \sum_f \bra{f} \Pro_i \Pro_V \ket{f} = \tr [\Pro_i \Pro_V]
\end{equation}
where $\Pro_V \equiv \sum_{h_V} \ket{h_V} \bra{h_V}$ is the projector onto localized 
states $\ket{h_V}$ and $\Pro_i = {\sf P}_{P J \lambda \Pi}$ defined in Sect.~2. 
In ref.~\cite{micro1}, we argued that $\Pro_V$ can be written, in its most general
form, in a field theoretical framework as a functional integral over eigenstates 
$\ket{\psi}$ of the (scalar charged) quantum field:
\begin{eqnarray}\label{pv1}
&&   \Pro_V = \int_V \mathcal{D}(\psi,\psi^\dagger) \ket{\psi,\psi^\dagger}
   \bra{\psi,\psi^\dagger} \nonumber \\
&& {\rm with} \qquad \ket{\psi,\psi^\dagger} \equiv 
   \otimes_{{\bf x}\in V} \ket{\psi({\bf x}),\psi^\dagger({\bf x}')} 
   \otimes_{{\bf x}\notin V}\ket{\psi({\bf x}),\psi^\dagger({\bf x}')}
\end{eqnarray}
where the field function $\psi({\bf x})$ is some arbitrary function outside the region 
$V$\footnote{We will use the same symbol $V$ to denote the system region and its volume.}. 
In the functional integration (\ref{pv1}), the index $V$ implies that integrated field 
variables are just those in $V$, i.e. 
${\mathcal D} \psi = \prod_{{\bf x}\in V} d \psi({\bf x})$. With this definition,
spurious terms depending on the degrees of freedom outside $V$ are introduced which must 
be subtracted ``by hand'' in a consistent way. This means that no contribution arising
from these degrees of freedom should be retained at the end of the calculation 
\cite{micro1}.

The quantum field theoretical approach was suggested in ref.~\cite{meaning} to be
essential for small volumes, because localized states are in principle a superposition
of states with different number of particles (meant as asymptotic states of the free 
field defined over the whole space). The approximation of assuming localized states
having a well-defined number of particles is expected to be a good one only when the
size of the region is much larger than Compton wavelength. Yet, the final expression
of the microcanonical partition function obtained in ref.~\cite{micro1} only differs
by that calculated with the aforementioned approximation by an immaterial constant factor,
after due subtraction of spurious terms. In this work, we will extend this result to
particles with spin, that is to non-scalar fields. 

As a first step, the MPF $\Omega$ is expressed as an expansion over all possible 
{\em channels}. By channel we mean a set of particle multiplicities for each species, 
that is a set of integers $N_1, \ldots ,N_K \equiv \Nj$ where $K$ is the total number
of species. In formula:
\begin{equation}\label{chanexp}
 \Omega = \sum_{\Nj} \Omega_{\Nj} \; .
\end{equation}
$\Omega_{\Nj}$ is defined as the {\em microcanonical channel weight} and can be 
calculated integrating on kinematical degrees of freedom. Writing $\ket{f} \equiv 
\ket{\Nj,\{ k\}}$ where by $\{ k \}$ we label the set of kinematical variables 
(momenta and spins) of the free Fock space state $\ket{f}$, the microcanonical 
weight of the channel $\Nj$ reads:
\begin{equation}\label{chanexp2}
 \Omega_{\Nj} = \sum_{\{ k \}} \omega_{ \Nj, \{ k \}} 
= \sum_{\{ k \}} \bra{ \Nj, \{ k \}} \Pro_i \Pro_V  \ket{\Nj,\{ k \}} \; .
\end{equation}
where $\omega_{ \Nj, \{ k \}}$ is defined as the {\em microcanonical state weight}.  
It is now useful to insert a resolution of the identity in $\omega_{ \Nj, \{ k \}}$
by using again the completeness of a set of Fock space states $\ket{ \Nj', \{ k'\}}$:
\begin{equation}\label{omega}
 \omega_{ \Nj, \{ k \}} \equiv \sum_{ \Nj', \{ k'\}} 
 \bra{\Nj, \{ k \}} \Pro_i \ket{\Nj', \{ k'\}} \bra{\Nj', \{ k'\}} \Pro_V 
 \ket{\Nj, \{ k \}} 
\end{equation}
and work out the matrix elements $\bra{f} \Pro_i \ket{f'}$ and $\bra{f'}\Pro_V \ket{f}$ 
separately. It is easy to realize that $\Pro_i$ in (\ref{omega}) cannot change the 
multiplicities of the ket $\ket{f'}$. This can be seen from its integral expression 
(\ref{final}): translation, rotations and reflections cannot change the number of 
particles of the state vector they are acting upon. Therefore, in the sum in 
eq.~(\ref{omega}) $N'=N$ and:
\begin{equation}\label{omega1}
 \omega_{ \Nj, \{ k \}} \equiv \sum_{\{ k'\}} 
 \bra{\Nj, \{ k \}} \Pro_i \ket{\Nj, \{ k'\}} \bra{\Nj, \{ k'\}} \Pro_V 
 \ket{\Nj, \{ k \}} 
\end{equation}

We will start our derivation of the MPF by considering the projector 
$\Pro_i\equiv {\sf P}_{P J \lambda}$ without parity fixing, which will be addressed 
in Sect.~10. According to eq.~(\ref{final}), the reduced projector 
${\sf P}_{P J \lambda}$ can be written as:
\begin{equation}\label{proji1}
 \Pro_{P J \lambda} = \delta^4( P - \hat{P}) (2J + 1) \int \d{\sf R} \; 
 D^{J}_{\lambda \lambda}({\sf R}^{-1}) \, \hat{{\sf R}} \, \; .
\end{equation}
We will first work out the channel with one particle and calculate $\Omega_{1}$.
This will serve as a basis for the calculation of the general microcanonical channel
weight in Sec.~5.

%****************************************************************************
\section{Single particle channel}
%****************************************************************************

Henceforth, we will stick to the notation of ref.~\cite{moussa}. Single particle states
are written as $\ket{[p],\sigma}$ and fulfill:
\begin{equation} \label{proji3}
 \hat{P}\ket{[p],\sigma} = p \ket{[p],\sigma} \qquad {\rm and} \qquad 
 \hat{S}_3(p)\ket{[p],\sigma} = \sigma \ket{[p],\sigma} 
\end{equation}
with normalization: 
\begin{equation}\label{norma}
 \braket{[p],\sigma}{[q],\tau} = \delta^3({\bf p}-{\bf q})\delta_{\sigma\tau}\;.
\end{equation}
With $[p]$ we denote the SL(2,C) (universal covering group of SO(1,3)$^{\uparrow}_+$)
matrix corresponding to the Lorentz transformation transforming $p_0=(m,{\bf 0})$ 
into $p=(\varepsilon,{\bf p})$. This is similar to the notation of eq.~(\ref{choi}), 
with the difference that now the notation is extended to SL(2,C) and expressions
like $[p]^\dagger$ are now meaningful. In the previous equations, square brackets in 
$\ket{[p],\tau}$ warns us that, strictly speaking, the corresponding state (and 
also the operator $\hat{S}_3(p)$) has been defined relatively to a non-unique 
set of axes $p,n_i(p);\; i=1, 2, 3$ (see Sect.~2), but this will not affect our 
calculation in any way \cite{moussa} and we will drop henceforth this notation 
letting:
\begin{equation}
\ket{[p],\sigma}=\ket{p,\sigma} \; .
\end{equation}
The transformation of a state $\ket{p,\sigma}$ under a general Lorentz transformation 
${\sf \Lambda}$ reads:
\begin{equation} \label{lorentztr}
\hat{{\sf \Lambda}}\ket{p,\sigma} = \sum_\tau \ket{{\sf \Lambda}p,\tau} 
D^S_{\tau \sigma}([{\sf \Lambda}p]^{-1}{\sf \Lambda} [p]) 
\sqrt{\frac{(\Lambda p)^0}{p^0}}\; .
\end{equation}
where $S$ is the spin of the particle, $({\sf \Lambda} p)^0$, $p^0$ are respectively 
the time component of ${\sf \Lambda} p$, $p$ and $[{\sf \Lambda}p]^{-1}{\sf \Lambda} 
[p]$ is the Wigner rotation. A special case of (\ref{lorentztr}) occurs when 
${\sf \Lambda}$ is a rotation:
\begin{equation} \label{proji4}
\hat{{\sf R}}\ket{p,\sigma} = \sum_\tau \ket{{\sf R}p,\tau} 
D^S_{\tau\sigma}([{\sf R}p]^{-1}{\sf R} [p]) \; .
\end{equation}
According to eq.~(\ref{omega1}), we have to evaluate:
\begin{equation}\label{omega1-2}
 \omega_{1,(p\sigma)} \equiv \sum_{p',\sigma'} \bra{p,\sigma} \Pro_i \ket{p',\sigma'} 
 \bra{p',\sigma'} \Pro_V \ket{p,\sigma} \; .
\end{equation}
By using the integral expansion of the projector $\Pro_{P J \lambda}$ in 
eq.~(\ref{proji1}), we can easily provide an integral expression of the first matrix 
element in eq.~(\ref{omega1-2}): 
\begin{eqnarray} \label{matrixsingle}
\bra{p,\sigma} \Pro_{P J \lambda} \ket{p',\sigma'}&=&\delta^4(P - p) (2J + 1) 
\int \d{\sf R} \; D^{J}_{\lambda  \lambda}({\sf R}^{-1}) \nonumber \\  
&\times& \delta^3({\sf R}{\bf p}'-{\bf p}) 
D^S_{\sigma\sigma'}([{\sf R}p']^{-1}{\sf R}[p']) \; .
\end{eqnarray}

Conversely, it is much more difficult to develop the expression of the second matrix
element $\bra{p,\sigma}\Pro_V\ket{p,\sigma'}$. For spinless particles, this requires
a functional method, described in detail in ref.~\cite{micro1}. To cope with this 
problem, we first need the expression of fields associated to particles with generic 
spin $S$. The free fields corresponding to the $2(2S+1)$ physical degrees of freedom 
of a particle and antiparticle with spin $S$ in Heisenberg representation can be 
written as \cite{moussa} (see also ref.~\cite{weinspin}):
\begin{eqnarray}\label{fields}
\!\!\!\!\!\!\!\!\!\!
\Psi_\tau(x)&=&\frac{1}{(2 \pi)^{\frac{3}{2}}} \sum_\tau 
 \int \frac{\d^3 {\rm p}}{\sqrt{2 \varepsilon}} 
 \; D^S_{\tau \sigma}([p]) a(p,\sigma) \e^{-i p \cdot x}+D^S_{\tau \sigma}([p] C^{-1})
 b^\dagger(p,\sigma) \e^{i p \cdot x} \nonumber \\
\!\!\!\!\!\!\!\!\!\!
\tilde{\Psi}_\tau(x)&=&\frac{1}{(2 \pi)^{\frac{3}{2}}} \sum_\tau \int \frac{\d^3 {\rm p}}
{\sqrt{2 \varepsilon}} \; D^S_{\tau\sigma}([p]^{\dagger -1})
a(p,\sigma) \e^{-i p \cdot x} + D^S_{\tau \sigma}([p]^{\dagger -1} C)
b^\dagger(p,\sigma) \e^{i p \cdot x} \nonumber \\
&&
\end{eqnarray} 
where $\varepsilon=\sqrt{{\rm p}^2+m^2}$ is the energy; $a,\;a^\dagger$ and 
$b,\;b^\dagger$ are respectively destruction and creation operators for particles 
and antiparticles; $D^S$ is the finite-dimensionsal representation of SL(2,C) 
of dimension $2S+1$ and the SL(2) matrix $C \equiv i \sigma_2$ ($\sigma_i$ 
being a Pauli matrix). The (anti)commutation relations of creation and destruction
operators in (\ref{fields}) are related to the normalization of states (\ref{norma})
and read:
\begin{equation}\label{commut}
[a(p,\sigma),a^\dagger(p',\sigma')]_{\pm}= [b(p,\sigma),b^\dagger(p',\sigma')]_{\pm}=
\delta^3({\bf p}-{\bf p}') \delta_{\sigma\sigma'} \; .
\end{equation}
Under a Lorentz (or SL(2,C)) transformation, the fields in eq.~(\ref{fields}) 
transform according to \cite{moussa,weinspin}:
\begin{eqnarray}\label{fieldtransf}
&&\hat{{\sf \Lambda}}\Psi_\tau(x)\hat{{\sf \Lambda}}^{-1}=
\sum_\sigma D^S_{\tau \sigma}({\sf \Lambda}^{-1})\Psi_\sigma({\sf \Lambda}\, x) 
 \nonumber \\
&&\hat{{\sf \Lambda}}\tilde{\Psi}_\tau(x)\hat{{\sf \Lambda}}^{-1}=
 \sum_\sigma D^S_{\tau \sigma}({\sf \Lambda})^{\dagger} 
 \tilde{\Psi}_\sigma({\sf \Lambda}\, x) \; .
\end{eqnarray}
where we have written them in the Heisenberg representation. Note that 
$D^S({\sf \Lambda})$ and $D^S({\sf \Lambda})^{\dagger -1}$ are two inequivalent
finite-dimensional non-unitary representation of the Lorentz group and correspond to 
those labelled as $(S,0)$ and $(0,S)$ respectively. As representations of SL(2,C) 
they have useful properties:
\begin{equation}\label{matricid}
  D^S(A^\dagger)=D^S(A)^\dagger \qquad D^S(A^T)=D^S(A)^T \; .
\end{equation} 
The matrix $C=i \sigma_2$ fulfills following relations:
\begin{equation}\label{cmatr}
 C=-C^{T}=C^{*} \qquad C^2=-1 \qquad CAC^{-1}A^{T} = (\det A) \, I \;\; \forall A 
 \in  {\rm SL(2,C)}  
\end{equation}
and plays a crucial role in the spin-statistics connection, as:
\begin{equation}\label{spinstat}
  D^S_{\tau \sigma}(C^2)=(-1)^{2S} \delta_{\tau \sigma} \; .
\end{equation}

The $2(2S+1)$ field degrees of freedom in eq.~(\ref{fields}) are needed to 
represent particle and antiparticle. Since 
$\tilde{\Psi}_\tau(x)=D^S_{\tau \sigma}(C)\Psi^{c\dagger}_\sigma(x)$; where 
$\Psi^c$ is the charge-conjugated field which is obtained from $\Psi$ by swapping 
$a \leftrightarrow b $ and $a^\dagger \leftrightarrow b^\dagger $ in the eq.~(\ref{fields}),
for a particle coinciding with its antiparticle, $\tilde{\Psi}_\tau(x) = 
D^S_{\tau \sigma}(C) \Psi^{\dagger}_\sigma(x)$, a constraint which effectively 
halves the field degrees of freedom. 

It is easily seen from (\ref{fields}) that for equal times $t=t'$ 
$[\Psi({\bf x}),\tilde\Psi({\bf x}')]_\pm=0$, where the sign $+$ or $-$ applies
to the fermion or boson case respectively, so it is possible to put together the 
fields $\Psi$ and $\tilde\Psi$ in one spinor $\Psis$ with $2(2S+1)$ independent 
(anti)commuting components:
\begin{equation}
 \Psis = {\Psi \choose \tilde{\Psi}} 
\end{equation}
and write the two fields in eq.~(\ref{fields}) in a compact way as:
\begin{equation}\label{fields2}
  \Psis(x) = \frac{1}{(2 \pi)^{\frac{3}{2}}} \sum_\sigma
  \int \frac{\d^3 {\rm p}}{\sqrt{2 \varepsilon}} \; U(p,\sigma) a(p,\sigma) 
  \e^{- i p \cdot x} + V(p,\sigma) b^\dagger(p,\sigma) \e^{i p \cdot x} 
\end{equation}
where $U(p,\sigma)$ and $V(p,\sigma)$ are $2(2S+1)$-dimensional spinors: 
\begin{equation}\label{spinors0}
 U(p,\sigma) = \left( \begin{array}{c} D^S_{1,\sigma}([p])  \\
 \vdots  \\ D^S_{2S+1,\sigma}([p])  \\ D^S_{1,\sigma}([p]^{\dagger -1})  \\
 \vdots  \\ D^S_{2S+1,\sigma}([p]^{\dagger -1}) \end{array} \right)
 \qquad
 V(p,\sigma) = \left( \begin{array}{c} D^S_{1,\sigma}([p]C^{-1})  \\
 \vdots  \\ D^S_{2S+1,\sigma}([p]C^{-1})  \\ D^S_{1,\sigma}([p]^{\dagger -1}C)  \\
 \vdots  \\ D^S_{2S+1,\sigma}([p]^{\dagger -1}C) \end{array} \right)
\end{equation}
It is also possible to write the field in an even more compact way defining the 
$2(2S+1)\times(2S+1)$ matrices $U(p)$ and $V(p)$ by clumping the spinors
$U(p,\sigma)$ and $V(p,\sigma)$ respectively:
\begin{equation}\label{spinors}
 U(p) = {D^S([p]) \choose D^S([p]^{\dagger -1})} \qquad
 V(p) = {D^S([p]C^{-1}) \choose D^S([p]^{\dagger -1}C)} 
\end{equation}
so that the eq.~(\ref{fields2}) can be recast as:
\begin{equation}\label{fields3}
  \Psis(x) = \frac{1}{(2 \pi)^{\frac{3}{2}}} 
  \int \frac{\d^3 {\rm p}}{\sqrt{2 \varepsilon}} \; \left( U(p) a(p) 
  \e^{-i p \cdot x} + V(p) b^\dagger(p) \e^{i p \cdot x} \right)
\end{equation}
$a(p)$ and $b^\dagger(p)$ being $2S+1$-dimensional column vectors of destruction
and creation operators. Now, let us define the $(2S+1)\times 2(2S+1)$
conjugate spinorial matrices:
\begin{equation}
 \bar U(p) = U^\dagger(p) \Gamma_0  \qquad \bar V(p) = V^\dagger(p) 
 \Gamma_0   
\end{equation}
$\Gamma_0$ being the $2(2S+1)\times2(2S+1)$ matrix:
$$
  \Gamma_0 = \left( \begin{array}{cc} 0 \; & \; {\rm I} \\ {\rm I} \; & 
 \; 0 \end{array} 
  \right) \; .
$$ 
The following relations apply:
\begin{eqnarray}\label{ortho}
&& \bar U(p) U(p) = (-1)^{2S} \bar V(p) V(p) = 2{\rm I}  \nonumber \\
&& \bar U(p) V(p) = \bar V(p) U(p) = 
 \left\{ \begin{array}{l} 2D^S(C) \;\; {\rm if} \, S \, {\rm even} \\ 0 
  \;\; {\rm if} \, S \, {\rm odd} \end{array} \right.
\end{eqnarray}
as well as:
\begin{eqnarray}\label{ortho2}
&&  U(p) \bar U(p) = \left( \begin{array}{cc} {\rm I} \; & \; D^S([p][p]^\dagger) \\
                   D^S([p][p]^\dagger)^{-1} \; & \; {\rm I} \end{array} \right)    
 \nonumber \\ 
&&  V(p) \bar V(p) = \left( \begin{array}{ll} (-1)^{2S}{\rm I} \; & \;  
        D^S([p][p]^\dagger) \\ D^S([p][p]^\dagger)^{-1}\; & \; (-1)^{2S}{\rm I} 
     \end{array} \right)
\end{eqnarray}
One can construct the conjugate field $\bar \Psis$:
\begin{equation}\label{conjuf}
  \bar\Psis(x) \equiv \Psis^\dagger(x) \Gamma_0 =
  \frac{1}{(2 \pi)^{\frac{3}{2}}} \int \frac{\d^3 {\rm p}}{\sqrt{2 \varepsilon}}\;
  \left( a^\dagger(p) \bar U(p) \e^{i p \cdot x} + b(p) \bar V(p) \e^{-i p \cdot x}
  \right) 
\end{equation}
The fields $\Psis$ and $\bar\Psis$ obey homogenous generalized Dirac equations
for spin $S$ \cite{moussa,weinspin}. Particularly, for $S=1/2$ it can be shown 
that $\sqrt(m) \Psis$ is the usual Dirac field and $\Psis$ obeys the familiar 
Dirac equations in the Weyl representation:
\begin{equation}\label{Dirac}
 i \gamma^{\mu} \partial_\mu \Psis - m \Psis = 0 \qquad \gamma^0 = \Gamma_0 \;\;
 \gamma^i = \left( \begin{array}{cc} 0 & -\sigma^i \\ \sigma^i & 0
            \end{array} \right)   \; .
\end{equation}

To proceed with the calculation of the matrix element 
$\bra{p,\sigma}\Pro_V\ket{p,\sigma'}$ we now have to express the creation and 
destruction operators as a function of the fields. We do this by using the 
fields in Schr\"odinger representation, i.e. with the fields at $t=0$:
\begin{eqnarray}\label{invert}
\bra{0} a(p,\sigma)&=& \bra{0} \; \frac{1}{2(2 \pi)^{\frac{3}{2}}}\sum_\tau 
\int \d^3 {\rm x} \; \e^{-i {\bf p} \cdot {\bf x}} \sqrt{2 \varepsilon}\,
\bar U(p)_{\sigma\tau}\Psis_{\tau}({\bf x}) \nonumber \\
a^\dagger(p,\sigma) \ket{0} &=& \frac{1}{2(2 \pi)^{\frac{3}{2}}}\sum_\tau 
\int \d^3 {\rm x}\; \e^{i {\bf p} \cdot {\bf x}} \sqrt{2 \varepsilon} \,
 \bar\Psis_\tau({\bf x}) U(p)_{\tau\sigma} \;\ket{0} \nonumber \\
 \bra{0} b(p,\sigma)&=& \bra{0} \; \frac{1}{2(2 \pi)^{\frac{3}{2}}}\sum_\tau 
\int \d^3 {\rm x} \; \e^{-i {\bf p} \cdot {\bf x}} \sqrt{2 \varepsilon}\, 
\bar{\Psis}_\tau({\bf x}) V(p)_{\tau\sigma} (-1)^{2S} \nonumber \\ 
 b^\dagger(p,\sigma)\ket{0} &=&\frac{1}{2(2 \pi)^{\frac{3}{2}}}\sum_\tau 
 \int \d^3 {\rm x}\; \e^{i {\bf p} \cdot {\bf x}} \sqrt{2 \varepsilon}\,
  \bar V(p)_{\sigma\tau} \Psis_\tau({\bf x}) (-1)^{2S} \; \ket{0} \; ;
\end{eqnarray}
these relations can be checked on the basis of (\ref{ortho}). Therefore, 
using (\ref{invert}) for particles:
\begin{eqnarray}\label{expand1}
&&\!\!\!\!\!\!\!\!\!\!
 \bra{p',\sigma'} \Pro_V \ket{p,\sigma} = \bra{0} a(p',\sigma') \Pro_V 
 a^\dagger(p,\sigma) \ket{0} \nonumber \\
&& \!\!\!\!\!\!\!\!\!\!
 = \frac{2\sqrt{\varepsilon \varepsilon'}}{4(2 \pi)^3}
 \int \d^3 {\rm x}' \int \d^3 {\rm x} \; 
 \e^{-i {\bf p}' \cdot {\bf x}'+i {\bf p}\cdot {\bf x}} \sum_{\tau,\tau'}
 \bar U(p')_{\sigma'\tau'} \bra{0}\Psis_{\tau'}({\bf x'}) \Pro_V \bar\Psis_{\tau}({\bf x})
 \ket{0} U(p)_{\tau\sigma}  \nonumber \\
&&
\end{eqnarray}
The difficult task in the above equation is to calculate the vacuum expectation 
value in presence of the projector $\Pro_V$. For the scalar case, we could cope
with this problem through a functional integration of the field degrees of freedom
in the region $V$ and dropping the terms involving the points outside the region
$V$ \cite{micro1}. However, the functional method developed for the scalar field
cannot be trivially carried over to the non-scalar case for a twofold reason:
firstly, even in the bosonic case, the fields $\Psis$ and $\bar\Psis$ do not commute 
at equal times and, as a consequence, writing an expansion of the operator $\Pro_V$ 
like in eq.~(\ref{pv1}) is not possible; secondly, one has to deal with fermionic 
degrees of freedom.

Yet we have some clues of how to work out (\ref{expand1}), getting inspiration from 
the scalar case. The presence of $\Pro_V$ means that eventually we have to restrict
all spacial integrations to the region $V$. Then, we will {\em assume} that:
\begin{equation}\label{keypart}
 \int_V \d^3 {\rm x}' \; \e^{-i {\bf p}'\cdot {\bf x}'} 
 \bra{0}\Psis_{\tau'}({\bf x'}) \Pro_V \bar\Psis_{\tau}({\bf x})\ket{0} = 
 \frac{\e^{-i {\bf p}'\cdot {\bf x}}}{2\varepsilon'} \,
 (U(p')\bar U(p'))_{\tau'\tau}\bra{0}\Pro_V\ket{0}
\end{equation}
and:
\begin{equation}\label{keyapart}
 \int_V \d^3 {\rm x}' \; \e^{-i {\bf p}'\cdot {\bf x}'} 
 \bra{0}\bar\Psis_{\tau'}({\bf x'}) \Pro_V \Psis_{\tau}({\bf x})\ket{0} = 
 \frac{\e^{-i {\bf p}'\cdot {\bf x}}}{2\varepsilon'} \,
 (V(p')\bar V(p'))_{\tau'\tau}\bra{0}\Pro_V\ket{0} \; .
\end{equation}
The justification of the formulae (\ref{keypart}) and (\ref{keyapart}) resides 
in two facts:
\begin{itemize}
\item{} they are consistent with the limit $\Pro_V \to {\sf I}$, as it can be
 readily checked from the explicit field expression and the normalization 
 condition (\ref{norma});
\item{} they have been proved for the scalar case \cite{micro1}.
\end{itemize}
If one accepts the conjecture expressed by the equations (\ref{keypart}) and 
(\ref{keyapart}) the eq.~(\ref{expand1}) turns into, by using (\ref{ortho}):
\begin{eqnarray}\label{expand2} 
  &&\bra{p',\sigma'} \Pro_V \ket{p,\sigma} \nonumber \\
  &=& \frac{\bra{0} \Pro_V \ket{0}}{4(2 \pi)^3} \sum_{\tau\tau'}
 \sqrt{\varepsilon \over \varepsilon'} \int_V \d^3 {\rm x} \; 
 \e^{-i {\bf x} \cdot ({\bf p}'-{\bf p})} \bar U(p')_{\sigma'\tau'} 
 (U(p')\bar U(p'))_{\tau'\tau} U(p)_{\tau\sigma}  \nonumber \\
 &=& \frac{1}{2} \, \sqrt{\varepsilon \over \varepsilon'} \, F_V ({\bf p}-{\bf p}') 
 (\bar U(p') U(p))_{\sigma'\sigma}\bra{0} \Pro_V \ket{0} \nonumber \\
 &=& \frac{1}{2} \, \sqrt{\varepsilon \over \varepsilon'} \, F_V ({\bf p}-{\bf p}') 
 \left( D^S([p']^{-1}[p])+D^S([p']^{\dagger}[p]^{\dagger-1}) \right)_{\sigma'\sigma}
 \bra{0} \Pro_V \ket{0}
\end{eqnarray}
where $F_V$ is a Fourier integral over the system region $V$:
\begin{equation}\label{fint}
F_V({\bf p}-{\bf p}') = \frac{1}{(2\pi)^3} 
\int_V \d^3 {\rm x} \; \e^{i {\bf x \cdot}({\bf p}-{\bf p}')} \; .
\end{equation}
For $p=p'$ and $\sigma=\sigma'$ the above expression reduces to $V/(2\pi)^3
\bra{0} \Pro_V \ket{0}$ which is the same we obtained for the scalar case in 
ref.~\cite{micro1}. It is straightforward to show that the same expression 
holds for antiparticles. 

We are now in a position to calculate the microcanonical state weight $\omega_1$. 
By plugging (\ref{matrixsingle}) and (\ref{expand2}) in the definition 
(\ref{omega1-2}): 
\begin{eqnarray}\label{omega1-3}
\!\!\!\!\!\!\!\!\!\!
&& \omega_1 = \nonumber \\
\!\!\!\!\!\!\!\!\!\!
&=& \delta^4(P - p) \sum_{\sigma'} \int \d^3{\rm p}' \;  (2J + 1) 
\int \d{\sf R} \; D^{J}_{\lambda  \lambda}({\sf R}^{-1}) 
\delta^3({\sf R}{\bf p}'-{\bf p}) D^S_{\sigma\sigma'}([{\sf R}p']^{-1}{\sf R}[p'])
 \nonumber \\
\!\!\!\!\!\!\!\!\!\!
&\times& \frac{1}{2}\sqrt{\varepsilon \over \varepsilon'} F_V ({\bf p}-{\bf p}') 
 \left( D^S([p']^{-1}[p])+D^S([p']^{\dagger}[p]^{\dagger-1} \right)_{\sigma'\sigma}
 \bra{0} \Pro_V \ket{0} \nonumber \\
\!\!\!\!\!\!\!\!\!\!
&=& \delta^4(P - p) \sum_{\sigma'} \int \d^3{\rm p}' \; (2J + 1) 
\int \d{\sf R} \; D^{J}_{\lambda  \lambda}({\sf R}^{-1}) 
\delta^3({\sf R}{\bf p}'-{\bf p}) D^S_{\sigma\sigma'}([p]^{-1}{\sf R}[p'])
 \nonumber \\
\!\!\!\!\!\!\!\!\!\!
&& \times \frac{1}{2} F_V ({\bf p}-{\bf p}') \left( D^S([p']^{-1}[p])+
  D^S([p']^{\dagger}[p]^{\dagger-1} \right)_{\sigma'\sigma}\bra{0} \Pro_V \ket{0}
\end{eqnarray}
Two matrix products are to be computed in the above equation. The first yields:
\begin{equation}\label{mp1}
 \sum_{\sigma'} D^S_{\sigma\sigma'}([p]^{-1}{\sf R}[p'])
 D^S_{\sigma'\sigma}([p']^{-1}[p])) = D^S([p]^{-1}{\sf R}[p])_{\sigma\sigma}
\end{equation}
To work out the second matrix product we observe that $[{\sf R}p']^{-1}{\sf R}[p']=
[p]^{-1}{\sf R}[p']$ is a Wigner {\em rotation}, hence it is unitary:
$$
 [p]^{-1}{\sf R}[p']=([p]^{-1}{\sf R}[p'])^{-1\dagger}=
 ([p']^{-1}{\sf R}^{-1}[p])^{\dagger}=([p]^{\dagger}{\sf R}[p']^{\dagger-1})
$$
thus:
\begin{eqnarray}\label{mp2}
\!\!\!\!\!\!\!\!\!\!
 && \sum_{\sigma'} D^S_{\sigma\sigma'}([p]^{-1}{\sf R}[p'])
 D^S_{\sigma'\sigma}([p']^\dagger [p]^{\dagger-1})= \sum_{\sigma'}
 D^S_{\sigma\sigma'}([p]^{\dagger}{\sf R}[p']^{\dagger-1})
 D^S_{\sigma'\sigma}([p']^\dagger [p]^{\dagger-1}) \nonumber \\
\!\!\!\!\!\!\!\!\!\!
 && = D^S_{\sigma\sigma}([p]^{\dagger}{\sf R}[p]^{\dagger-1})
\end{eqnarray}
By using (\ref{mp1}) and (\ref{mp2}) the eq.~(\ref{omega1-3}) becomes:
\begin{eqnarray}\label{omega1-4}
&& \omega_1 =  \delta^4(P - p) \int \d^3{\rm p}' \; (2J + 1) 
 \int \d{\sf R} \; D^{J}_{\lambda  \lambda}({\sf R}^{-1}) 
 \delta^3({\sf R}{\bf p}'-{\bf p}) \nonumber \\
&& \times \frac{1}{2} F_V ({\bf p}-{\bf p}') \left(D^S([p]^{-1}{\sf R}[p])
 +D^S([p]^{\dagger}{\sf R}[p]^{\dagger-1})\right)_{\sigma\sigma}
 \bra{0} \Pro_V \ket{0}
\end{eqnarray}
This is the final expression of the single-particle channel. We note in passing
that, with the same method, the more general matrix element 
$\bra{p, \sigma'}\Pro_i \Pro_V \ket{p,\sigma}$ could have been calculated, yielding:
\begin{eqnarray}\label{omega1-5}
&& \bra{p ,\sigma'}\Pro_i \Pro_V \ket{p,\sigma}= \delta^4(P - p)
 \int \d^3{\rm p}' \;  (2J + 1) 
 \int \d{\sf R} \; D^{J}_{\lambda  \lambda}({\sf R}^{-1}) 
 \delta^3({\sf R}{\bf p}'-{\bf p}) \nonumber \\
&& \times \frac{1}{2} F_V ({\bf p}-{\bf p}') \left(D^S([p]^{-1}{\sf R}[p])
 +D^S([p]^{\dagger}{\sf R}[p]^{\dagger-1})\right)_{\sigma'\sigma}
 \bra{0} \Pro_V \ket{0}
\end{eqnarray}
%

%****************************************************************************
\section{Multiparticle channels}
%****************************************************************************

We first tackle the problem of a channel with $N$ particles of different species.
Writing the Fock state $\ket{N, \{ k\} }$ in eq.~(\ref{omega1}) as:
$$
\ket{f} = \ket{N, \{ k\} } = \ket{ N ,\{ p_{\parti}\},\{ \sigma_{\parti} \}}
$$
where $\{ p_{\parti} \}$; $\{ \sigma_{\parti} \}$ label the set of four-momenta 
and helicities of the $N$ particles, the microcanonical state weight in 
eq.~(\ref{omega1}) becomes:
\begin{eqnarray}\label{distint}
\omega_f &=&  \sum_{\{ \sigma'_{\parti} \}} \left[\prod_{\parti=1}^{N} \int \d^3 
 {\rm p}'_{\parti} \right]  \bra{N ,\{ p_{\parti} \},\{ \sigma_{\parti} \}} 
 \Pro_{P J \lambda} \ket{N , \{ p'_{\parti} \},\{\sigma'_{\parti} \}} \nonumber \\
  &\times& \bra{N,\{ p'_{\parti} \},\{\sigma'_{\parti} \}} \Pro_V 
  \ket{N,\{p_{\parti}\},\{\sigma_{\parti}\}} \; .
\end{eqnarray}
Both matrix elements of $\Pro_{P J \lambda}$ and $\Pro_V$ can be calculated by a
straightforward generalization of the single-particle channel because, as 
particles are distinct, creation and destruction operators, as well as field operators,
commute with each other and the problem is fully factorizable. Therefore:
\begin{eqnarray}\label{matr1}
&&\bra{ N, \{ p_\parti \},\{ \sigma_\parti \}} \Pro_{P J \lambda} 
 \ket{ N, \{ p'_\parti \},\{ \sigma'_\parti \}}=
 \delta^4 \left( P - \sum_{\parti=1}^{N} p_\parti \right)\nonumber \\  
 &&\times \; (2J + 1) \int \d{\sf R} \; D^{J}_{\lambda\lambda} ({\sf R}^{-1}) 
  \left[ \prod_{\parti=1}^{N} 
  D^{S_\parti}_{\sigma_\parti \sigma'_\parti}([{\sf R}p'_\parti]^{-1}{\sf R}[p'_\parti]) 
 \delta^3({\sf R}{\bf p}'_\parti-{\bf p}_\parti) \right] 
\end{eqnarray}
and:
\begin{eqnarray}\label{matr2} 
  && \bra{ N, \{ p_\parti  \},\{ \sigma'_\parti \}} \Pro_V
  \ket{ N, \{ p'_\parti  \},\{ \sigma_\parti \}} =  \bra{0} \Pro_V \ket{0}\nonumber \\
 && \times \prod_{\parti=1}^{N} \frac{1}{2}\sqrt{\varepsilon_\parti \over \varepsilon'_\parti} 
  F_V ({\bf p}_\parti-{\bf p}'_\parti) \left(  D^S([p'_\parti]^{-1}[p_\parti])+
  D^S([p'_\parti]^{\dagger}[p_\parti])^{\dagger-1} \right)_{\sigma'_\parti\sigma_\parti}
  \; .
\end{eqnarray}
Putting (\ref{matr1}) and (\ref{matr2}) into (\ref{distint}) and working out 
the matrix products, like in eqs.~(\ref{mp1}),(\ref{mp2}) we get:
\begin{eqnarray}\label{distint2}
\!\!\!\!\!\!\!\!\!\!
&& \omega_f= \bra{0} \Pro_V \ket{0} \, (2J + 1) \int \d{\sf R} \; 
  D^{J}_{\lambda  \lambda}({\sf R}^{-1}) \left[\prod_{\parti=1}^{N}\int 
  \d^3 {\rm p}'_{\parti} \; \delta^3({\sf R}{\bf p}_{\parti}'-{\bf p}_{\parti}) \right. 
  \nonumber \\
\!\!\!\!\!\!\!\!\!\!
&&\times \left. \frac{1}{2} 
 \left(D^S([p_\parti]^{-1}{\sf R}[p_\parti])+D^S([p_\parti]^{\dagger}{\sf R}
 [p_\parti]^{\dagger-1})\right)_{\sigma_\parti\sigma_\parti} \!\!\!
  F_V({\bf p}_{\parti}-{\bf p'}_{\parti}) \right] 
 \delta^4 \left( P - \sum_{\parti=1}^{N} p_{\parti} \right) \nonumber \\
\!\!\!\!\!\!\!\!\!\!
&&
\end{eqnarray}
The integration over momenta ${\rm p}'_{\parti}$ can now be performed. Since
${\bf p}'_{\parti}={\sf R}^{-1}{\bf p}_{\parti}$:
\begin{eqnarray}\label{distint3}
\!\!\!\!\!\!\!\!\! \omega_f &=& \bra{0} \Pro_V \ket{0} \,
  \delta^4 \left(P - \sum_{\parti=1}^{N} p_\parti \right)
  (2J + 1) \int \d{\sf R} \; D^{J}_{\lambda\lambda}({\sf R}^{-1})
  \nonumber \\ 
\!\!\!\!\!\!\!\!\! &\times& \left[\prod_{\parti=1}^{N}\frac{1}{2} 
 \left(D^S([p_\parti]^{-1}{\sf R}[p_\parti])+D^S([p_\parti]^{\dagger}{\sf R}
 [p_\parti]^{\dagger-1})\right)_{\sigma_\parti\sigma_\parti} 
  F_V({\bf p}_{\parti}-{\sf R}^{-1}{\bf p}_{\parti}) \right]  \; .
\end{eqnarray}
which is the final expression of the microcanonical state weight for $N$ different
particles. Integrating (\ref{distint3}) over momenta $p_\parti$ and summing 
over spin projections $\sigma_\parti$, one gets the microcanonical channel 
weight in a simple form: 
\begin{eqnarray}\label{chanweight}
\Omega_{N}&=& (2J + 1) \int \d{\sf R} \; D^{J}_{\lambda\lambda}({\sf R}^{-1}) 
\left[ \prod_{\parti=1}^{N} \int \d^3 {\rm p}_{\parti} \; \tr D^{S_\parti}({\sf R})
 \right. \nonumber \\  
 &\times& \left. 
 F_V({\bf p}_{\parti}-{\sf R}^{-1}{\bf p}_{\parti}) \right]
\delta^4 \left( P - \sum_{\parti=1}^{N} p_{\parti} \right) \bra{0}\Pro_V\ket{0} \; .
\end{eqnarray}
where we took advantage of cyclicity property of the trace. It is interesting 
to note that Wigner rotations and momentum-dependent matrices in the matrix element 
of $\Pro_V$ (\ref{matr2}) combine so as to leave, once the sum over spin projections
is carried out, a trace of a simple SU(2) rotation. 

We will now deal with the more complicated case of identical particles.

%====================================================================
\subsection{Identical particles}
%====================================================================

Developing the matrix element $\bra{f'} \Pro_V \ket{f}$ for a multiparticle state
with identical particles is far more difficult than for different particle species
because the associated fields do not commute and factorization does not hold; 
this has been shown for the scalar field in our previous work \cite{micro1}. 
Nevertheless, we showed that, keeping only terms involving field degrees of freedom
in the region $V$, the result is the same one would get in a multi-particle 
non-relativistic quantum mechanical framework enforcing symmetrization or antisymmetrization 
for bosons or fermions respectively. Although the conclusion was fair, the proof
involved lenghty calculations which would become even lenghtier for particles with 
spin. Therefore, we will not tackle the proof here and simply assume that it 
holds for particles with spin: {\it a general multiparticle state can be treated 
as a state with different particles, with (anti)symmetrization for identical particles}.
Accepting this conjecture allows us to work on the fixed-number multiparticle tensor 
space instead of the original Fock space. 

Let $\rho_j$ be a permutation of the integers $1,\ldots, N_j$, $\chi(\rho_j)$ its 
parity and $b_j=$\,0, 1 for bosons and fermions respectively. The generic state 
$\ket{f}$ in the multiparticle tensor space can be written as:
\begin{equation}\label{genstate}
\ket{f} = \sum_{ \{ \rho_j \}} \left[ \prod_{j=1}^{K} \frac{ \chi(\rho_j)^{b_j}}
{\sqrt{N_j !}}\right] \prod_{j=1}^{K} \ket{ N_j , \{ p_{ \rho_{j}(\parti_j) }  \} , 
\{\sigma_{ \rho_{j}(\parti_j) }  \}  } \; .
\end{equation}
where $K$ is the total number of particle species and $\{ p_{\parti_j} \}$, 
$\{\sigma_{\parti_j} \}$ the set of momenta and spin projections of particles of 
species $j$ in $\ket{f}$. Also let $\rhov$ be the set $(\rho_1, \ldots, \rho_K)$ of 
permutations of integers $(1, \ldots, N_1), \ldots,(1, \ldots, N_K)$ and let 
$N=\sum_{j=1}^k N_j$. Let us also introduce a shorthand $\ket{f_{\rhov}}$ for 
the ket:
\begin{equation}\label{shorthand}
\ket{f_{\rhov}} \equiv \prod_{j=1}^{K} \ket{ N_j , \{ p_{ \rho_{j}(\parti_j) }  
\} , \{ \sigma_{ \rho_{j}(\parti_j) }  \}  } 
\end{equation}
which is the multiparticle tensor formed by exchanging particle indices by the 
permutations $\rhov$. By using this notation, the state $\ket{f}$ in (\ref{genstate})
can be rewritten as:
\begin{equation}\label{genstate2} 
 \ket{f} = \sum_{ \rhov } \left[ \prod_{j=1}^{k} \frac{ \chi(\rho_j)^{b_j}}
 {\sqrt{N_j !}}\right] \; \ket{f_{\rhov}} \;
\end{equation}
and the microcanonical state weight $\omega_f$:
\begin{equation}\label{omegaid}
 \omega_f= \sum_{ \etav } \sum_{ \rhov} \left[ \prod_{j=1}^{K} 
 \frac{\chi(\rho_j \eta_j)^{b_j}}{N_j!} \right] 
 \bra{f_{\etav}}\Pro_i \Pro_V \ket{f_{\rhov}}
\end{equation}
which can be simplified, redefining dummy particle indices, into:
\begin{equation}\label{omegaid2}
 \omega_f= \sum_{ \rhov } \left[ \prod_{j=1}^{k} \chi(\rho_j)^{b_j} \right] 
 \bra{f_{\iotav}} \Pro_i \Pro_V \ket{f_{\rhov}}
\end{equation}
where $\iotav$ is the identical permutation.

We first consider, for sake of simplicity, a multiparticle state with $N$ identical 
particles. Eq.~(\ref{omegaid2}) reads, by taking $\Pro_i \equiv \Pro_{P J \lambda}$:
\begin{eqnarray}\label{omegan}
\omega_f &=& \sum_{\rho } \chi(\rho ) ^{b} \bra{f_{\iota} }
\Pro_{P J \lambda} \; \Pro_V \ket{ f_{\rho} } \nonumber \\ 
&=& \sum_{\rho } \chi(\rho ) ^{b} \bra{ N, \{ p_\parti  \},\{ \sigma_\parti \} }
\Pro_{P J \lambda} \; \Pro_V \ket{N,\{p_{\rho(\parti)}\},\{\sigma_{\rho(\parti)}\}}  
\end{eqnarray}
Like for eq.~(\ref{omega}), we insert a resolution of the identity between $\Pro_i$ and 
$\Pro_V$ leading to:
\begin{eqnarray}\label{omegan2}
\omega_f&=&\sum_{\rho } \chi(\rho )^{b}\left[\prod_{\parti=1}^{N} \sum_{\sigma'_\parti} 
\int \d^3 {\rm p}'_\parti\right] \; \bra{ N, \{ p_\parti  \},\{ \sigma_\parti \}}  
 \Pro_{P J \lambda} \ket{ N, \{ p'_\parti  \},\{ \sigma'_\parti \}} \nonumber \\  
&\times& \bra{ N, \{ p'_\parti  \},\{ \sigma'_\parti \} } \Pro_V \ket{ N,
\{ p_{\rho(\parti)}\},\{ \sigma_{\rho(\parti)} \}} \; .
\end{eqnarray}
Both matrix elements of $\Pro_V$ and $\Pro_{P J \lambda}$ can be evaluated, in the 
multiparticle tensor space, the same way as for distinct particles. By using 
eqs.~(\ref{matr1}) and (\ref{matr2}): 
\begin{eqnarray}\label{omegan3}
\!\!\!\!\!\!\!\!\!\!\!\!
&& \omega_f=\bra{0} \Pro_V \ket{0} \delta^4 \left( P - \sum_{\parti=1}^{N} p_\parti \right)
 \left[\prod_{\parti=1}^{N} \sum_{\sigma'_\parti} 
 \int \d^3 {\rm p}'_\parti\right](2J + 1) \nonumber \\
\!\!\!\!\!\!\!\!\!\!\!\! &&\times \sum_{\rho } \chi(\rho )^{b}\int \d{\sf R} \; 
 D^{J}_{\lambda  \lambda}({\sf R}^{-1})   
\left[ \prod_{\parti=1}^{N} \delta^3({\sf R}{\bf p}'_\parti-{\bf p}_\parti)  
D^S_{\lambda_\parti  \lambda'_\parti}([{\sf R}p'_\parti]^{-1}{\sf R}[p'_\parti])
\right. \nonumber  \\
\!\!\!\!\!\!\!\!\!\!\!\!
 &&\times \left.\sqrt{\frac{\varepsilon_{\rho(\parti)}}{\varepsilon'_\parti}} 
 F_V({\bf p}_{\rho(\parti)}-{\bf p'}_\parti)\frac{1}{2} \left( 
 D^S([p'_\parti]^{-1}[p_{\rho(\parti)}])+ D^S([p'_\parti]^\dagger[p_{\rho(\parti)}]^
 {\dagger-1}) \right)_{\sigma'_\parti \sigma_{\rho(\parti)}} \right] \; .
\end{eqnarray}
If we now carry out the integration over momenta $p'_i$ and sum over all spin
projections $\sigma'_\parti$ we get:
\begin{eqnarray}\label{omegan4}
 \omega_f&=& \delta^4 \left( P - \sum_{\parti=1}^N p_\parti \right) 
 (2J + 1) \int \d{\sf R}\; D^{J}_{\lambda\lambda} ({\sf R}^{-1}) \sum_{\rho} \chi(\rho)^{b} \,   
 \nonumber \\ 
&\times& 
 \left[ \prod_{\parti=1}^{N} \frac{1}{2} \left( D^S([p_\parti]^{-1}{\sf R}[p_{\rho(\parti)}])+ 
 D^S([p_\parti]^\dagger{\sf R}[p_{\rho(\parti)}]^{\dagger-1})\right)_{\sigma_\parti
 \sigma_{\rho(\parti)}} \right. \nonumber \\
&\times& \left. 
 \sqrt{\frac{\varepsilon_{\rho(\parti)}}{\varepsilon_\parti}} \, 
  F_V({\bf p}_{\rho(\parti)}-{\sf R}^{-1}{\bf p}_\parti)\right] \bra{0}\Pro_V\ket{0} \; .
\end{eqnarray}
Indeed, the factor $\prod_{\parti=1}^N \sqrt{\varepsilon_{\rho(\parti)}/\varepsilon_\parti}$
can be dropped from eq.~(\ref{omegan4}) because numerator and denominator are altogether
equal being $\rho$ a permutation. 

The microcanonical channel weight of $N$ identical particles can be calculated by 
integrating $\omega_f$ in~(\ref{omegan4}):
\begin{eqnarray}\label{Omegan}
\!\!\!\!\!\!\!\!\!\!
 && \Omega_{N}=\bra{0}\Pro_V\ket{0}(2J + 1) \int \d{\sf R} \; 
 D^{J}_{\lambda\lambda}({\sf R}^{-1}) 
\sum_{\sigma_1,\ldots,\sigma_N} \sum_\rho \frac{\chi(\rho)^b}{N!} 
 \left[ \prod_{\parti=1}^{N} \int \d^3 {\rm p}_{\parti} \right. \nonumber \\
\!\!\!\!\!\!\!\!\!\!
&&\times \left. \frac{1}{2} \left( D^S([p_\parti]^{-1}{\sf R}[p_{\rho(\parti)}]) + 
 D^S([p_\parti]^\dagger{\sf R}[p_{\rho(\parti)}]^{\dagger-1})\right)_{\sigma_\parti
 \sigma_{\rho(\parti)}} \!\!\!
 F_V({\bf p}_{\rho(\parti)}- {\sf R}^{-1}{\bf p}_\parti) \right] \nonumber \\
\!\!\!\!\!\!\!\!\!\!
&&\times \delta^4 \left( P - \sum_{\parti=1}^N p_\parti \right)  \; 
\end{eqnarray}
where the factor $1/N!$ has been introduced in order avoid multiple counting of 
momenta and spin projections.

The above formula can be further developed in the limit of large volumes, when:
$$
  F_V({\bf p}_{\rho(\parti)}- {\sf R}^{-1}{\bf p}_\parti) 
  \underset{V \to \infty}{\longrightarrow}
 \delta^3({\bf p}_{\rho(\parti)}- {\sf R}^{-1}{\bf p}_\parti) 
$$ 
In this case, the two matrices of the irreducible representation of SL(2,C) in
the formula (\ref{Omegan}) become equal. To show this, one first notes that the
 Dirac's delta implies $p_{\rho(\parti)}={\sf R}^{-1}p_\parti$, hence:
\begin{eqnarray}
&& [p_\parti]^{-1}{\sf R}[p_{\rho(\parti)}] \rightarrow 
   [p_\parti]^{-1}{\sf R}[{\sf R}^{-1}p_\parti]  \nonumber \\
&& [p_\parti]^\dagger{\sf R}[p_{\rho(\parti)}]^{\dagger-1} \rightarrow 
   [p_\parti]^\dagger{\sf R}[{\sf R}^{-1}p_\parti]^{\dagger-1}
\end{eqnarray}
The right hand side of the first equation is manifestly a Wigner rotation, hence
it is unitary and equal to the right hand side of the second equation. This 
implies that in the large volume limit:
\begin{equation}\label{reduction}
\frac{1}{2} \left( D^S([p_\parti]^{-1}{\sf R}[p_{\rho(\parti)}]) + 
 D^S([p_\parti]^\dagger{\sf R}[p_{\rho(\parti)}]^{\dagger-1})\right) 
 \underset{V \to \infty}{\longrightarrow}
 D^S([p_\parti]^{-1}{\sf R}[p_{\rho(\parti)}])
\end{equation}
We observe that the same equation applies in the non-relativistic limit, where
the matrix $[p]$ is unitary, since the associated Lorentz boost in eq.~(\ref{choi}) 
can be approximated with the identity. 

An useful development of the formula in the thermodynamic limit can be achieved 
by decomposing permutations into the product of irreducible cyclic permutations, 
that is writing: 
\begin{equation}\label{cycdec}
\rho = c_1 \ldots c_H\, .
\end{equation}
Instead of working out the most general case, we will focus on an example which 
will show more clearly how to take advantage of the above decomposition; the 
generalization will then be straightforward. Let us then consider a special case, 
with $5$ particles and the permutation:
\begin{equation}\label{permu}
(1,2,3,4,5) \qquad \longrightarrow \qquad (3,1,2,5,4)
\end{equation}
which can be decomposed in two cyclic permutations: 
\begin{equation}\label{eqn:permu2}
c_1=(1,2,3) \longrightarrow (3,1,2)\qquad {\rm and} 
\qquad c_2=(4,5)\longrightarrow(5,4) \;.
\end{equation}
If we now write down the corresponding matrices product in eq.~(\ref{Omegan}), 
by using the reduction (\ref{reduction}) we have:
\begin{eqnarray*}\label{permu3}
&&\prod_{\parti=1}^{5} D^{S}_{\sigma_{\parti}\sigma_{\rho(\parti)}}
 ([p_{\parti}]^{-1}{\sf R}[p_{\rho(\parti)}])= 
 D^{S}_{\sigma_{4} \sigma_{5}}([p_4]^{-1}{\sf R}[p_{5}])D^{S}_{\sigma_{5} 
 \sigma_{4}}([p_5]^{-1}{\sf R}[p_{4}]) \nonumber \\
&&\qquad \qquad \times D^{S}_{\sigma_{1} \sigma_{3}}([p_1]^{-1}{\sf R}[p_{3}])
 D^{S}_{\sigma_{3} \sigma_{2}}([p_3]^{-1}{\sf R}[p_{2}])
 D^{S}_{\sigma_{2} \sigma_{1}}([p_2]^{-1}{\sf R}[p_{1}])
\end{eqnarray*}
where the first two factors on the right hand side correspond to the cyclic 
permutation $c_2$, while the other three to $c_1$. Summing then over spin
projections $\sigma_\parti$, we end up with the simple and nice result:
\begin{eqnarray}\label{traces}
 \sum_{\{ \sigma_{\parti} \}} \prod_{\parti=1}^{5} D^{S}_{\sigma_{\parti} 
 \sigma_{\rho(\parti)}}([p_{\parti}]^{-1}{\sf R}[p_{\rho(\parti)}])
 &=& 
 \tr D^{S}([p_4]^{-1}{\sf R}^2[p_{4}]) \tr D^{S}([p_1]^{-1}{\sf R}^3 [p_{1}])   
 \nonumber \\  
&=& \tr D^{S}({\sf R}^2) \tr D^{S}({\sf R}^3 ) 
\end{eqnarray}
In general, for each permutation $\rho$ and its cyclic decomposition, if we let 
$l$ be the number of integers in each cyclic permutation and $h_{l}$ the number 
of cyclic permutations with $l$ elements in $\rho$ such that $\sum_{l=1}^\infty 
l h_{l} = N$ \footnote{The set of integers $h_1,\ldots,h_{N} \equiv \{h_{l}\}$,
is usually defined as a {\em partition} of the integer $N$ in the multiplicity 
representation.}, one has:
\begin{equation}\label{generaldec}
 \sum_{\{ \sigma \}} 
 \prod_{\parti=1}^{N} D^{S}_{\sigma_{\parti}\sigma_{\rho(\parti)}}
 ([p_{\parti}]^{-1}{\sf R}[p_{\rho(\parti)}]) = 
 \prod_{l=1}^{N} \left[ \tr D^{S}({\sf R}^{l})\right]^{h_{l}(\rho)}
\end{equation}
Thus, by using eq.~(\ref{reduction}) and its consequence eq.~(\ref{generaldec})
in the large volume limit the microcanonical channel weight (\ref{Omegan}) turns 
into:
\begin{eqnarray}\label{Omegan2}
 \Omega_{N}&=& (2J + 1)  \int \d{\sf R} \; 
 D^{J}_{\lambda  \lambda}({\sf R}^{-1}) \sum_{\rho} \frac{\chi(\rho)^{b}}{N!} 
 \left[ \prod_{\parti=1}^{N} \int \d^3 {\rm p}_{\parti} \right. \nonumber \\
&\times& \left. F_V({\bf p}_{\rho(\parti)}- {\sf R}^{-1}{\bf p}_{\parti}) 
\left[ \tr D^{S}({\sf R}^{\parti}) \right]^{h_{\parti}(\rho)} \right] 
\delta^4 \left(P - \sum_{\parti=1}^{N} p_\parti \right) \bra{0} \Pro_V \ket{0}
 \, .
\end{eqnarray}
an expression which also applies to the non-relativistic limit. On the other
hand, if $V$ is not large and we have a full relativistic regime, the two 
$D^S$ matrices in eq.~(\ref{Omegan}) are no longer equal and the summation 
over polarizations gives rise to a very involved expression. The general 
(\ref{Omegan}) and the special case (\ref{Omegan2}) can be easily extended to 
a multi-species ideal gas. They read:
\begin{eqnarray}\label{General1}
 \Omega_{\Nj} &=& (2J + 1) \int \d{\sf R} \; D^{J}_{\lambda\lambda}({\sf R}^{-1}) 
 \sum_{\sigma_1,\ldots,\sigma_N} \sum_{\rhov} \left[ \prod_{j=1}^K 
 \frac{\chi(\rho_j)^{b_j} }{N_j!} \prod_{\parti_j=1}^{N_j} \right.  
 \int \d^3{\rm p}_{n_j} \nonumber \\  
  &\times&\frac{1}{2}\left( D^{S_j}([p_{\parti_j}]^{-1}{\sf R}[p_{\rho(\parti_j)}])+ 
 D^{S_j}([p_{\parti_j}]^{\dagger}{\sf R}[p_{\rho(\parti_j)}]^{\dagger-1}) \right)
_{\sigma_{\parti_j}\sigma_{\rho(\parti_j)} } \nonumber \\
 &\times& \left. F_V({\bf p}_{\rho(\parti_j)}- {\sf R}^{-1}{\bf p}_{\parti_j}) \right]
 \delta^4 \left(P - \sum_{\parti=1}^{N} p_\parti \right) \bra{0} \Pro_V \ket{0}
\end{eqnarray}
and:
\begin{eqnarray}\label{General2}
\!\!\!\!\!\!\!\!\!\!
&& \Omega_{\Nj}= (2J + 1) \int \d{\sf R} \; D^{J}_{\lambda\lambda}({\sf R}^{-1}) 
 \sum_{\rhov} \left[ \prod_{j=1}^K \frac{\chi(\rho_j)^{b_j} }{N_j!} 
 \prod_{\parti_j=1}^{N_j} \right. \int \d^3{\rm p}_{n_j} \nonumber \\
\!\!\!\!\!\!\!\!\!\!  
&&\times\left. F_V({\bf p}_{\rho(\parti_j)}- {\sf R}^{-1}{\bf p}_{\parti_j}) 
   \left[\tr D^{S_j}({\sf R}^{n_j}) \right]^{h_{n_j}(\rho_j)} \right]
 \delta^4 \left(P - \sum_{\parti=1}^{N} p_\parti \right) \bra{0} \Pro_V \ket{0}
\end{eqnarray}
respectively. In the eq.~(\ref{General2}) $\sum_{n_j=1}^{N_j} n_j h_j(\rho_j) = N_j$.

Finally, we observe that the eq.~(\ref{General2}) could be further developed in 
the large volume limit so as to obtain a microcanonical channel weight which is 
resummable analytically yielding a compact integral expression of the MPF, just like 
in the non-relativistic case without angular momentum conservation \cite{bf1}. 

In the large volume limit expression (\ref{General2}) there appear traces of rotations. 
Since the trace is the invariant over group conjugacy classes, it turns out that 
all SU(2) transformations of the same angle $\psi$ around an axis $\hat{{\bf n}}$ 
(which labels the different members of the same conjugacy class) have the same 
trace. It is therefore convenient to use the axis-angle parametrization for the 
integration measure over the SU(2) group:
\begin{eqnarray}\label{su2measure}
 \int \d{\sf R} &=& \frac{1}{16 \pi^2} \int_{0}^{\pi} \d \theta \int_{0}^{2 \pi} 
 \d \phi \int_{0}^{4 \pi} \d \psi \; 2 \sin \theta  \sin^2 \frac{\psi}{2} \\ \nonumber
 &=& \frac{1}{16 \pi^2} \int \d\Omega_{\hat{{\bf n}}} \int_{0}^{4 \pi} \d \psi \; 
 2 \sin^2 \frac{\psi}{2} 
\end{eqnarray}
where $(\theta,\phi)$ are the polar and azimuthal angles defining the axis $\hat{{\bf n}}$.
With this parametrization, the trace of a rotation ${\sf R}_{\hat{{\bf n}}}(\psi)$ 
reads \cite{lgroup}:
\begin{equation}\label{trace}
\tr D^{S}\left({\sf R}_{\hat{{\bf n}}}(\psi) \right) = 
\frac{\sin[\left(S+\frac{1}{2} \right) \psi]}{ \sin \frac{\psi}{2}} \; .
\end{equation}
Furthermore, since:
$$
 {\sf R}_{\hat{{\bf n}}}(\psi_1){\sf R}_{\hat{{\bf n}}}(\psi_2)=
 {\sf R}_{\hat{{\bf n}}}(\psi_1+\psi_2) \Rightarrow 
{\sf R}_{\hat{{\bf n}}}(\psi)^l={\sf R}_{\hat{{\bf n}}}(l\psi)
$$ 
the trace in eq.~(\ref{General2}) simply becomes:
\begin{equation}
 \tr D^{S}({\sf R}_{\hat{{\bf n}}}(\psi)^{n}) =
 \frac{\sin[(S+\frac{1}{2}){n}\psi]}{\sin(\frac{n\psi}{2})} \; .
\end{equation}
%

%**************************************************************************
\section{Spherical region}
%**************************************************************************

The eq.~(\ref{General1}) is the most general microcanonical weight of a channel 
$\Nj$ with the enforcement of energy-momentum and angular momentum conservation; 
eq.~(\ref{General2}) is an approximate expression of (\ref{General1}) which applies
for very large volumes. Summing (\ref{General1}) or (\ref{General2}) over all 
possible channels yields the full microcanonical partition function.

If the system region has spherical symmetry, those expressions can be further
simplified because the whole SU(2) group integrand in all the equations is independent 
of the axis $\hat{{\bf n}}$ (see end of Sect.~5). The proof of this statement proceeds 
in two steps: first, one observes that $\Omega_{\Nj}$ as a whole does not depend on 
the polarization index $\lambda$ for a spherical region, so that one can rewrite
it summing over $\lambda$ and dividing by $2J+1$. In the second step, one shows that 
the overall SU(2) integrand does not depend on the axis $\hat{{\bf n}}$. In formulae:
\begin{eqnarray}\label{integrand}
\Omega_{\Nj} &=& (2J+1) \int \d {\sf R} \; D^{J}({\sf R}^{-1})_{\lambda\lambda} 
  I({\sf R}) = \sum_\lambda \int \d {\sf R} \; D^{J}({\sf R}^{-1})_{\lambda\lambda} 
  I({\sf R}) \nonumber \\
 &\equiv& \int \d {\sf R} \; \tr D^J({\sf R}^{-1}) I({\sf R}) 
\end{eqnarray}
where neither $\tr D^J({\sf R}^{-1})$ nor $I({\sf R})$ depend on the axis polar
coordinates. 

Let us prove the first part of our statement by writing the projector 
$\Pro_{P J \lambda}$ for a system at rest (i.e. with ${\bf P}=0$) as:
\begin{equation}\label{sphepro}
\Pro_{P J \lambda} = \ket{M,J,\lambda}\bra{M,J,\lambda}
\end{equation}
(where $M$ is the mass of the system) and the channel weight in (\ref{chanexp2}) 
as:
\begin{eqnarray}\label{sphechan}
 \Omega_{\Nj} &=& \sum_{\{k\}} \braket{ \Nj,\{k\}}{M,J,\lambda} 
 \bra{M,J,\lambda}  \Pro_V  \ket{\Nj,\{k\}} \\ \nonumber 
 &=& \bra{M,J,\lambda}  \Pro_V \left( \sum_{\{k\}} \ket{\Nj,\{k\}} \bra{ \Nj,\{k\}} 
 \right) \ket{M,J,\lambda} \; .
\end{eqnarray}
Evidently, for spherically symmetric systems the condition:
\begin{equation}\label{sphecomm}
  [\Pro_V, \hat{{\sf R }}]=0
\end{equation}
must hold, because $\Pro_V$ (see e.g. eq.~(\ref{pv1})) involves the sum of all 
possible field configurations within a spherically symmetric region and the 
representation of a rotation on the Hilbert space transforms $\Pro_V$ into itself. 
Likewise, the operator:
\begin{equation}\label{prochan}
\Pro_{\Nj} \equiv \sum_{\{ p \}} \ket{\Nj,\{ p \}} \bra{ \Nj,\{ p \}}
\end{equation}
is also invariant under rotations since it is a sum over all possible kinematical  
configurations of free states. Thus: 
\begin{equation}\label{sphecomm2}
[\Pro_{\Nj},\hat{{\sf R }}]=0  
\end{equation}
and, because of Wigner-Eckart theorem:
\begin{equation}\label{we}
  \Omega_{\Nj}= \bra{M,J,\lambda}\Pro_V\Pro_{\Nj}\ket{M,J,\lambda} =  
  \bra{M,J}|\Pro_V\Pro_{\Nj} |\ket{M,J} \; ,
\end{equation}
as $\Pro_V \Pro_{\Nj}$ commutes with rotation operators according to (\ref{sphecomm}) 
and (\ref{sphecomm2}). Therefore, the channel weight is independent of $\lambda$. One 
can then sum over spin projections $\lambda$ the matrix element in all expressions of the 
microcanonical channel weight, and divide by $(2J+1)$, i.e.:
\begin{equation}\label{sumtrace}
\frac{1}{(2J+1)}\sum_{\lambda} D^{J}_{\lambda \lambda}
({\sf R}_{\hat{{\bf n}}}^{-1}(\psi)) = 
\frac{1}{(2J+1)}\tr D^{J}({\sf R}_{\hat{{\bf n}}}^{-1}(\psi))=
\frac{1}{(2J+1)}\frac{\sin \left(J+\frac{1}{2} \right) \psi}
{ \sin \frac{\psi}{2}}
\end{equation}
where we used $\tr [{\sf R}_{\hat{{\bf n}}}^{-1}(\psi)]= 
\tr[{\sf R}_{\hat{{\bf n}}}(-\psi)]=\tr[{\sf R}_{\hat{{\bf n}}}(\psi)]$. Thereby,
the dependence on $\hat {\bf n}$ owing to the matrix $D^J({\sf R}_{\hat{{\bf n}}}
(\psi))$ is removed.

The second step of the proof is to show that the function $I({\sf R})$ in the
eq.~(\ref{integrand}) is also independent of the axis polar coordinates. By comparing 
eq.~(\ref{integrand}) with the integral expression of $\Pro_{J\lambda}$, i.e. 
$$
\Pro_{J\lambda} = (2J+1) \int \d {\sf R} D^{J}({\sf R}^{-1})_{\lambda\lambda} \hat{\sf R}
$$
and comparing (\ref{sphechan}) with (\ref{integrand}) one has:
\begin{equation}\label{integrand2}
 I({\sf R}) = \sum_{\{ k\}} \bra{ \Nj,\{k\}} \delta^4(P-\hat P) \hat{\sf R} \Pro_V
 \ket{ \Nj,\{k\}} = \tr [\delta^4(P-\hat P) \hat{\sf R} \Pro_V \Pro_{\Nj}]
\end{equation}
using definition (\ref{prochan}). Let us now replace in the above equation the
rotation ${\sf R}$ with axis $\hat{\bf n}$ with that around the rotated axis ${\sf O}
\hat{\bf n}$, which is ${\sf O}{\sf R}{\sf O}^{-1}$. The rigthmost term in 
eq.~(\ref{integrand2}) turns into:   
\begin{eqnarray}\label{tracefin}
&& \tr [\delta^4(P-\hat P) \hat{\sf 0} \hat{\sf R} \hat{\sf 0}^{-1} \Pro_V \Pro_{\Nj}]=
  \tr [\hat{\sf 0} \delta^4(P-\hat P) \hat{\sf R} \Pro_V \Pro_{\Nj} \hat{\sf 0}^{-1}]
\nonumber \\
&& = \tr [\delta^4(P-\hat P) \hat{\sf R} \Pro_V \Pro_{\Nj}]
\end{eqnarray}
where we used eqs.~(\ref{sphecomm}),(\ref{sphecomm2}) and took advantage of the
commutation between $\delta^4(P-\hat P)$ and rotations if ${\bf P}=0$. The eq.~
(\ref{tracefin}) proves our statement, namely the function $I({\sf R})$ is independent
of the rotation axis and so is the whole group integrand function in eq.~(\ref{integrand}).

In conclusion, for a spherically symmetric region, one can integrate away the solid 
angle $\Omega_{\hat{{\bf n}}}$ and choose an arbitrary rotation axis, e.g. $\hat{\bf k}$, 
in all expressions of the microcanonical channel weight $\Omega_{\Nj}$. The integration 
measure of the SU(2) group (\ref{su2measure}) in the axis-angle parametrization 
effectively reduces to:
\begin{equation}\label{redmeas}
 \int \d{\sf R} = \frac{1}{16 \pi^2} \int \d\Omega_{\hat{{\bf n}}} \d \psi \; 2 \sin^2 
 \frac{\psi}{2} \rightarrow \frac{1}{4 \pi} \int_{0}^{4 \pi} \d \psi \; 2 \sin^2 
 \frac{\psi}{2} \; . 
\end{equation}
Particularly, the two equations (\ref{General1}) and (\ref{General2}) for a spherical
region become:  
\begin{eqnarray}\label{spherical1}
\!\!\!\!\!\!\!\!\!\!
 \Omega_{\Nj} &=& \frac{1}{2 \pi} \int_0^{4\pi} \d \psi \; \sin {\psi \over 2}
 \sin \left(J+\frac{1}{2}\right) \psi  
 \sum_{\sigma_1,\ldots,\sigma_N} \sum_{\rhov} \left[ \prod_{j=1}^K 
 \frac{\chi(\rho_j)^{b_j} }{N_j!} \prod_{\parti_j=1}^{N_j} \right.  
 \int \d^3{\rm p}_{n_j} \nonumber \\  
\!\!\!\!\!\!\!\!\!\!
  &\times&\frac{1}{2}\left( D^{S_j}([p_{\parti_j}]^{-1}{\sf R}_3(\psi)[p_{\rho(\parti_j)}])+ 
 D^{S_j}([p_{\parti_j}]^{\dagger}{\sf R}_3(\psi)[p_{\rho(\parti_j)}]^{\dagger-1}) \right)
_{\sigma_{\parti_j}\sigma_{\rho(\parti_j)} } \nonumber \\
 &\times& \left. F_V({\bf p}_{\rho(\parti_j)}- {\sf R}_3(\psi)^{-1}{\bf p}_{\parti_j}) \right]
 \delta^4 \left(P - \sum_{\parti=1}^{N} p_\parti \right) \bra{0} \Pro_V \ket{0}
\end{eqnarray}
and:
\begin{eqnarray}\label{spherical2}
\!\!\!\!\!\!\!\!\!\!
&& \Omega_{\Nj} = \frac{1}{2 \pi} \int_0^{4\pi} \d \psi \; \sin {\psi \over 2}
 \sin \left(J+\frac{1}{2}\right) \psi 
 \sum_{\rhov} \left[ \prod_{j=1}^K \frac{\chi(\rho_j)^{b_j} }{N_j!} 
 \prod_{\parti_j=1}^{N_j} \right. \int \d^3{\rm p}_{n_j} \nonumber \\ 
\!\!\!\!\!\!\!\!\!\! 
&&\times\left. F_V({\bf p}_{\rho(\parti_j)}- {\sf R}^{-1}_3(\psi){\bf p}_{\parti_j}) 
\left[\frac{\sin(S_j+\frac{1}{2}){\parti_j}\psi}{\sin(\frac{\parti_j\psi}{2})}\right]
^{h_{\parti_j}(\rho_j)}\right]
 \delta^4 \left(P - \sum_{\parti=1}^{N} p_\parti \right) \bra{0} \Pro_V \ket{0}
\nonumber \\
\!\!\!\!\!\!\!\!\!\! &&
\end{eqnarray}
%

%***********************************************************************
\section{Partial wave expansion}
%***********************************************************************

So far, we have been using a plane wave expansion to calculate the microcanonical 
channel weight. One may wonder whether an equivalent formula could be obtained 
using single particle states with definite angular momentum. In this section, 
we will show that this is the case. 
 
Instead of starting from scratch pur calculation by expressing the microcanonical 
state weight (\ref{omega}) with kets $\ket{j m S \sigma}$ or $\ket{j m l S}$ (in 
the $lS$-coupling), we will work out the formula eq.~(\ref{General1}) expanding 
plane waves into partial waves. For the sake of simplicity, we restrict to the simple 
case of Boltzmann statistics, corresponding to retain only the permutation 
identity in the sum in (\ref{General1}), yielding:
\begin{eqnarray}\label{Boltzmann}
\!\!\!\!\!\!\!\!\!\!
 \Omega_{\Nj} &=& (2J + 1) \int \d {\sf R} \; D^{J}_{\lambda\lambda}({\sf R}^{-1}) 
 \left[ \prod_{j=1}^K \frac{1}{N_j!} \prod_{\parti_j=1}^{N_j} 
  \int \d^3 {\rm p}_{n_j} \right. \nonumber \\  
\!\!\!\!\!\!\!\!\!\!
&\times& \left. \frac{1}{(2\pi)^3} \int \d^3 {\rm x} 
 \e^{i {\bf x}\cdot({\bf p}_{\parti_j}-{\sf R}^{-1}{\bf p}_{\parti_j})} \, 
 \tr D^{S_j}({\sf R}) \right]
 \delta^4 \left(P - \sum_{\parti=1}^{N} p_\parti \right) \bra{0} \Pro_V \ket{0} 
 \nonumber \\
\!\!\!\!\!\!\!\!\!\!
&=& (2J + 1) \int \d {\sf R} \; D^{J}_{\lambda\lambda}({\sf R}^{-1}) 
 \left[ \prod_{j=1}^K \frac{1}{N_j!} \right] \left[ \prod_{\parti=1}^{N} \right. 
  \int \d^3 {\rm p}_{n} \nonumber \\ 
\!\!\!\!\!\!\!\!\!\! 
&\times& \left. \frac{1}{(2\pi)^3} \int \d^3 {\rm x} 
 \e^{i {\bf x}\cdot({\bf p}_{\parti}-{\sf R}^{-1}{\bf p}_{\parti})} \, 
 \tr D^{S_n}({\sf R}) \right]
 \delta^4 \left(P - \sum_{\parti=1}^{N} p_\parti \right) \bra{0} \Pro_V \ket{0} 
\end{eqnarray}
where we used the explicit form of integrals (\ref{fint}) and performed the sum over
polarization states in eq.~(\ref{General1}) for the simple case of only identical
permutation. In the second equality we dropped different indices for different
particle species and used, for simplicity, a common index $n$ for particles in the
channels.

The exponentials in Fourier integrals in eq.~(\ref{Boltzmann}) can now be expanded 
into partial waves according to the well known formulae:
\begin{eqnarray}\label{pwave_exp2}
&&\e^{i{\bf x} \cdot {\bf p}_{\parti}} = \sum_{l_\parti=0}^{\infty} 
 \sum_{m_\parti=-l_\parti}^{l_\parti} i^{\,l_\parti} 4 \pi 
 j_{l_\parti}({\rm p}_{\parti} {\rm x}) Y_{l_\parti m_\parti}(\hat{{\bf x}}) 
 Y^*_{l_\parti m_\parti} (\hat{{\bf p}}_{\parti}) \\ \nonumber
&&\e^{-i{\bf x} \cdot {\sf R}{\bf p}_{\parti}} = \sum_{l'_\parti=0}^{\infty} 
\sum_{m'_\parti=-l'_\parti}^{l'_\parti} (-i)^{\,l'_\parti} 4 \pi j_{l'_\parti}
 ({\rm p}_{\parti} {\rm x}) Y^*_{l'_\parti m'_\parti} (\hat{{\bf x}}) 
 Y_{l'_\parti m'_\parti} ({\sf R}\hat{{\bf p}}_{\parti})
\end{eqnarray}
where $j_l$ are spherical Bessel functions and $Y$ are spherical harmonics; 
$\hat{{\bf x}}$ and $\hat{{\bf p}}$ are the unit vectors of ${\bf x}$ and ${\bf p}$
respectively. Consequently, Fourier integrals can be rewritten as:
\begin{eqnarray}\label{pwave_exp3}
&&\frac{1}{(2 \pi)^3}\int_V\d^3 {\rm x}\; \e^{i{\rm x} \cdot 
({\bf p}_{\parti}-{\sf R}{\bf p}_{\parti})}=
(4 \pi)^2 \!\!\!\! 
 \sum_{l_\parti,m_\parti;\; l'_\parti,m'\parti} i^{\,l_\parti}(-i)^{\,l'_\parti} 
 Y^*_{l_\parti m_\parti} (\hat{{\bf p}}_{\parti}) Y_{l'_\parti m'_\parti}
 ({\sf R}\hat{{\bf p}}_{\parti}) \nonumber \\  
&&\times \frac{1}{(2 \pi)^3}\int_V\d^3 {\rm x}\; j_{l_\parti}({\rm p}_{\parti}
{\rm x})j_{l'_\parti}({\rm p}_{\parti} {\rm x}) Y_{l_\parti m_\parti} (\hat{{\bf x}}) 
Y^*_{l'_\parti m'_\parti} (\hat{{\bf x}}) \; .
\end{eqnarray}
Switching to spherical coordinates $\d^3 {\rm x} \rightarrow {\rm x}^2 \, 
\d{\rm x}\, \d \Omega_{\hat{\bf x}}$, the angular integration can be made at once
yielding:
\begin{equation}\label{pwave_exp4}
\int \d \Omega_{\hat{\bf x}} Y_{lm} (\hat{{\bf x}})Y^*_{l'm'} (\hat{{\bf x}})= 
\delta_{l  l'} \delta_{m  m'} \; ,
\end{equation}
hence eq.~(\ref{pwave_exp3}) turns into:
\begin{eqnarray}\label{pwave_exp5}
&&  \frac{1}{(2 \pi)^3}\int_V\d^3 {\rm x}\; \e^{i{\rm x} \cdot 
({\bf p}_{\parti}-{\sf R}{\bf p}_{\parti})} \nonumber \\
&& =\frac{(4 \pi)^2}{(2 \pi)^3}\sum_{l_\parti,m_\parti} 
 Y^*_{l_\parti m_\parti} (\hat{{\bf p}}_{\parti})
 Y_{l_\parti m_\parti} ({\sf R}\hat{{\bf p}}_{\parti})
\int_V\d {\rm x}\; {\rm x}^2 j^2_l({\rm p}_{\parti} {\rm x}) \; .
\end{eqnarray}
In eq.~(\ref{Boltzmann}), for each particle, a factor $\tr D^{S}({\sf R})$ 
appears. By taking advantage of the relation:
\begin{equation}
 Y_{lm} ({\sf R}\hat{{\bf p}})=
 \sum_{m'} Y_{lm'}(\hat{{\bf p}})\; D^{l}_{m' m}({\sf R})^{*} 
\end{equation}
the product between the trace and $Y_{lm} ({\sf R}\hat{{\bf p}}_{\parti})$
can be worked out as:
\begin{eqnarray}\label{pwave_exp7}
\!\!\!\!\!\!\!\!\!\!
&& \tr D^{S_\parti}({\sf R}) \sum_{m'_\parti} Y_{l_\parti m'_\parti} 
 (\hat{{\bf p}}_{\parti})\; D^{l_\parti}_{m_\parti m'_\parti}({\sf R}) \nonumber \\
\!\!\!\!\!\!\!\!\!\!
&&=\sum_{m'_\parti,\sigma_{\parti}}Y_{l_\parti m'_\parti}(\hat{{\bf p}}_{\parti})\; 
D^{S_{\parti}}_{\sigma_{\parti}\sigma_{\parti}}({\sf R}) 
 D^{l_\parti}_{m_\parti m'_\parti}({\sf R}) \nonumber \\
\!\!\!\!\!\!\!\!\!\!
&& = \sum_{m'_\parti,\sigma_{\parti}}Y_{l_\parti m'_\parti}(\hat{{\bf p}}_{\parti})\; 
\bra{S_{\parti},\sigma_{\parti};l_\parti,m'_\parti} \hat{\sf R}
 \ket{S_{\parti},\sigma_{\parti};l_\parti,m_\parti} \nonumber  \\
\!\!\!\!\!\!\!\!\!\! 
&&=\sum_{\stackrel{J_\parti,\mu_\parti,\mu'_\parti}{m'_\parti,\sigma_{\parti}}}
 Y_{l_\parti m'_\parti}(\hat{{\bf p}}_{\parti})\; \bra{\sigma_{\parti}, m'_\parti} 
 S_{\parti} \, l_\parti \ket{J_\parti, \mu_\parti}\bra{J_\parti,\mu'_\parti} 
 S_{\parti} \, l_\parti \ket{\sigma_{\parti}, m_\parti} 
 D^{J_\parti}_{\mu_\parti \mu'_\parti}({\sf R})
\end{eqnarray}
where $\bra{m_1,m_2}j_1,j_2\ket{j,m}$ is a shorthand for the Clebsch-Gordan 
coefficient \\ 
$\braket{j_1,m_1;j_2,m_2}{j,m,j_1,j_2}$. In the last step, the rotation $\hat{\sf R}$ 
has been expanded over a complete set of vectors according to:
\begin{equation}\label{pwave_exp8}
 \hat{\sf R} =\sum_{J,\mu}\sum_{J',\mu'} \ket{J,\mu}\bra{J,\mu} \hat{\sf R} 
 \ket{J',\mu'}\bra{J',\mu'} 
 = \sum_{J,\mu} \sum_{J',\mu'} \ket{J,\mu} \bra{J',\mu'} \; 
 D^{J}_{\mu \mu'}({\sf R}) \delta_{J J'} \; .
\end{equation}
For each particle in the channel, a factor like (\ref{pwave_exp7}) appears. 
The SU(2) group integral in eq.~(\ref{Boltzmann}) can now be calculated collecting
all factors depending on ${\sf R}$:
\begin{eqnarray}\label{pwave_exp9}
&&(2J+1) \int \d {\sf R} \; D^J_{\lambda \lambda}({\sf R}^{-1}) 
D^{J_1}_{\mu_1 \mu'_1}({\sf R}) \ldots D^{J_N}_{\mu_N  \mu'_N}({\sf R}) \nonumber \\
&=& (2J+1) \bra{J_1,\mu_1 ; \ldots ; J_N,\mu_N} \int \d {\sf R} \; 
 D^J_{\lambda \lambda}({\sf R}^{-1}) \, {\hat{\sf R}} \ket{J_1,\mu'_1;\ldots ;J_N,\mu'_N} 
 \nonumber \\
&=& \bra{J_1,\mu_1; \ldots; J_N,\mu_N}{\sf P}_{J\lambda}\ket{J_1,\mu'_1; \ldots; J_N,\mu'_N} 
 \nonumber \\
&=& \braket{J_1,\mu_1; \ldots; J_N,\mu_N}{J, \lambda}\braket{J, \lambda}{J_1,\mu'_1; 
 \ldots; J_N,\mu'_N}
\end{eqnarray}
Collecting previous results, eqs~.(\ref{pwave_exp2})-(\ref{pwave_exp9}), the 
microcanonical channel weight (\ref{Boltzmann}) can be rewritten as:
\begin{eqnarray}\label{Boltzmann2}
&&\Omega_{\Nj}= \left[\prod_{j=1}^{k}\frac{1}{N_j!}\right] 
 \left[ \prod_{\parti=1}^{N} \int \d^3 {\rm p}_{\parti} \right] 
 \delta^4 \left(P - \sum_{\parti=1}^{N} p_\parti \right) \nonumber   \\
&&\times \sum_{\stackrel{J_\parti, \mu_\parti, \mu'_\parti}{l_\parti,m'_\parti, 
 \sigma_\parti, m_\parti}}\braket{J_1,\mu_1; \ldots; J_N,\mu_N}{J,\lambda}
\braket{J,\lambda} {J_1,\mu'_1; \ldots; J_N,\mu'_N}   \nonumber \\
&&\times 
\prod_{\parti=1}^{N} \frac{2}{\pi} \int_V\d {\rm x}_\parti\; {\rm x}_\parti^2 \; 
j^2_{l_\parti}({\rm p}_{\parti} {\rm x}_{\parti}) Y^*_{l_\parti m_\parti} 
(\hat{{\bf p}}_{\parti})Y_{l_\parti m'_\parti}(\hat{{\bf p}}_{\parti})  
\bra{\sigma_{\parti}, m'_{\parti}} S_{\parti} \, l_{\parti} \ket{J_{\parti}, 
\mu_{\parti}}  \nonumber\\
&&\times \bra{J_{\parti}, \mu'_{\parti}} S_{\parti} \, l_{\parti} 
\ket{\sigma_{\parti}, m_{\parti}} 
\end{eqnarray}
This expression finally gives a clear physical meaning to our manipulations: the 
microcanonical 
channel weight of the channel is obtained by summing over all possible single-particle 
states $\ket{p_n, l_n, m_n, S_n, \sigma_n}$, $l_n$ being the orbital angular momentum, 
$S_n$ the spin angular momentum, $m_n$ and $\sigma_n$ their respective third 
components; and projecting these states first on states with total particle angular 
momentum $J_n$ and then all of them onto a state with total angular momentum $J$ 
and third component $\lambda$. 

A considerable simplification occurs if, in eq.~(\ref{Boltzmann2}) momentum conservation 
constraint is dropped. In this case the $\delta^4$ reduces to 
$\delta \left(M - \sum_{\parti=1}^{N} \varepsilon_\parti \right)$ and the
integration over the angular coordinates of three-momenta can be done at once
yielding: 
\begin{equation}
\int \d \Omega_{\hat{\bf p}_\parti} Y_{l_\parti m_\parti}^*(\hat{{\bf p}}_\parti) 
 Y_{l_\parti m'_\parti} (\hat{{\bf p}}_\parti) = \delta_{m_\parti  m'_\parti} \; .
\end{equation}
Thereby, the eq.~(\ref{Boltzmann2}) reduces to, using the unitarity of Clebsch-Gordan
coefficients:
\begin{eqnarray}\label{ridotta}
&&\Omega_{\Nj}= \left[\prod_{j=1}^{k}\frac{1}{N_j!}\right]  
\left[ \prod_{\parti=1}^{N} \int \d {\rm p}_{\parti} {\rm p}^2_{\parti} \right]  
\delta \left(M - \sum_{\parti=1}^{N} \varepsilon_\parti \right) \nonumber \\
&&\times \sum_{J_\parti, \mu_\parti, l_\parti} 
\braket{J_1,\mu_1; \ldots; J_N,\mu_N}{J,\lambda}\braket{J,\lambda}
{J_1,\mu'_1; \ldots; J_N,\mu'_N}  \\ \nonumber 
&&\times 
\prod_{\parti=1}^{N} \frac{2}{\pi} \int_V\d {\rm x}_\parti\; {\rm x}_\parti^2 \; 
j^2_{l_\parti}({\rm p}_{\parti} {\rm x}_{\parti})  \; .
\end{eqnarray}
Let us now restore momentum conservation. In the rest frame of the system where:
\begin{displaymath}
\delta^4 \left(P - \sum_{\parti=1}^{N} p_\parti \right)=
\delta \left(M - \sum_{\parti=1}^{N} \varepsilon_\parti \right)
\delta^3 \left(\sum_{\parti=1}^{N} {\bf p}_\parti \right) 
\end{displaymath}
the $\delta^3$ is expand into spherical harmonics:
\begin{eqnarray}\label{pwave_exp11}
&&\delta^3 \left(\sum_{\parti=1}^{N} {\bf p}_\parti \right)= \frac{1}{(2 \pi)^3} 
\int \d^3 {\rm X} \; \e^{i \sum_{\parti=1}^N {\bf p}_\parti \cdot {\bf X}} = 
\frac{1}{(2 \pi)^3}\int \d^3 {\rm X} \; \prod_{\parti=1}^N 
\e^{i {\bf p}_\parti \cdot {\bf X}} \\ \nonumber
&=&\frac{1}{(2 \pi)^3}\int \d^3 {\rm X} \; 
\prod_{\parti=1}^N \sum_{L_\parti,k_\parti} \;i^{L_\parti} 4 \pi \; 
j_{L_\parti} ({\rm p}_\parti {\rm X}) Y^*_{L_\parti,k_\parti}
(\hat{\bf p}_\parti) Y_{L_\parti,k_\parti }(\hat{\bf X}) \; .
\end{eqnarray}
where $\hat{\bf X}$ is the unit vector of ${\bf X}$. 
Plugging eq.~(\ref{pwave_exp11}) into (\ref{Boltzmann}), we are left with an integration
of the product of three spherical harmonics having the same versor $\hat{\bf p}_\parti$ 
as argument. Switching to polar coordinates:
\begin{displaymath}
\d^3 {\rm p}_\parti \longrightarrow  \d {\rm p}_\parti\,{\rm p}^2_\parti  
\d \Omega_{\hat{\bf p}_\parti} \qquad {\rm and} \qquad \d^3 {\rm X} \longrightarrow  
\d {\rm X}\,{\rm X}^2  \d \Omega_{\hat{\bf X}}
\end{displaymath}
we have~\cite{sakurai}:
\begin{eqnarray}\label{pwave_exp12}
&&\int \d \Omega_{\hat{\bf p}_\parti} \; 
Y^*_{l_\parti m_\parti} (\hat{{\bf p}}_{\parti})
Y_{l_\parti m'_\parti}(\hat{{\bf p}}_{\parti})
Y^*_{L_\parti,k_\parti }(\hat{\bf p}_\parti) \\ \nonumber 
&=& \sqrt{\frac{2 L_\parti+1}{4 \pi}} \bra{0,0}l_\parti L_\parti \ket{l_\parti, 0}
\bra{m_\parti, k_\parti}l_\parti L_\parti \ket{l_\parti, m'_\parti} \; .
\end{eqnarray}
Furthermore, by recalling the definition of spherical harmonic:
\begin{equation}\label{pwave_exp13}
 \prod_{\parti=1}^N Y_{L_\parti,k_\parti }(\hat{\bf X}) = \prod_{\parti=1}^N 
 \sqrt{\frac{2 L_\parti+1}{4 \pi}} D^{L_\parti}_{k_\parti 0}(\varphi,\theta,0)^{*}
\end{equation}
where $\theta$ and $\varphi$ are, respectively, the polar and azimuthal angles of the
unit vector $\hat{\bf X}$, we can solve the angular integration in 
$\Omega_{\hat{\bf X}}$:
\begin{eqnarray}\label{pwave_exp14}
&& \int \d \Omega_{\hat{\bf X}} \; \prod_{\parti=1}^N \sqrt{\frac{2 L_\parti+1}
{4 \pi}} D^{L_\parti}_{k_\parti 0}(\varphi,\theta,0)^{*}  \nonumber \\
&=&\int \d \Omega_{\hat{\bf X}} \; \frac{1}{2 \pi}
\int_{0}^{2 \pi} \d \psi \prod_{\parti=1}^N \sqrt{\frac{2 L_\parti+1}{4 \pi}} 
D^{L_\parti}_{k_\parti 0}(\varphi,\theta,\psi)^{*} \nonumber \\
&=&
\prod_{\parti=1}^N \sqrt{\frac{2 L_\parti+1}{4 \pi}} 4 \pi \int \d {\sf R} 
\prod_{\parti=1}^N D^{L_\parti}_{0  k_\parti}({\sf R}^{-1}) \nonumber \\
&=&\prod_{\parti=1}^N \sqrt{\frac{2 L_\parti+1}{4 \pi}} 4 \pi 
\braket{L_1,0; \ldots;L_N,0}{0,0}\braket{0,0}{L_1,k_1; \ldots;L_N,k_N}
\end{eqnarray}
where we have used the unitarity of ${\sf R}$ and the last equality is a special
case of eq.~(\ref{pwave_exp9}).

Finally, the microcanonical channel weight reads:
\begin{eqnarray}\label{Boltzmann3}
&&\Omega_{\Nj}= \left[\prod_{j=1}^{k}\frac{1}{N_j!}\right] 
 \int_0^\infty \d {\rm X} \; {\rm X}^2  \left[\prod_{\parti=1}^{N} 
 \int \d {\rm p}_{\parti} \; {\rm p}^2_\parti \right]
 \delta \left(M - \sum_{\parti=1}^{N} \varepsilon_\parti \right) \nonumber \\
&&\times \sum_{\stackrel{J_\parti, \mu_\parti, \mu'_\parti, L_\parti, 
 k_\parti}{l_\parti, m'_\parti, \sigma_\parti, m_\parti}} \Bigg\{ \prod_{\parti=1}^{N} 
 \frac{2 L_\parti+1}{4 \pi} j_{L_\parti} ({\rm p}_\parti {\rm X})  
 \frac{2}{\pi} \int_V\d {\rm x}_\parti\; {\rm x}_\parti^2 \; 
 j^2_{l_\parti}({\rm p}_{\parti} {\rm x}_{\parti}) 
 \bra{0,0}l_\parti L_\parti \ket{l_\parti, 0} \nonumber  \\
&&\times
 \bra{m_\parti, k_\parti}l_\parti L_\parti 
 \ket{l_\parti, m'_\parti} \bra{\sigma_{\parti}, m'_{\parti}} S_{\parti} \, l_{\parti} 
 \ket{J_{\parti}, \mu_{\parti}}\bra{J_{\parti}, \mu'_{\parti}}  
 S_{\parti} \, l_{\parti} \ket{\sigma_{\parti}, m_{\parti}} 
 \Bigg\} \nonumber\\ 
&&\times  
 \braket{J_1,\mu_1; \ldots; J_N,\mu_N}{J,\lambda}\braket{J,\lambda}
 {J_1,\mu'_1; \ldots; J_N,\mu'_N} \nonumber \\
&&\times
 \braket{L_1,0; \ldots;L_N,0}{0,0}\braket{0,0}{L_1,k_1; \ldots;L_N,k_N} \; .
 \end{eqnarray}
In the above expression, only $N+1$ integrations are left instead of $3N$, but 
many series appear which can make the numerical calculation as cumbersome as in the
plane wave expansion (\ref{General1}).

%***************************************************************************
\section{Classical limit}
%***************************************************************************

We now show that the microcanonical channel weight we have calculated
in the quantum case reduces to what is expected for a fully classical system. 
The classical limit of eq.~(\ref{Boltzmann}) (we will not consider the quantum
statistics case) can be obtained by reintroducing $\hslash$ and then taking the
limit $\hslash \to 0$. Let us then rewrite eq.~(\ref{Boltzmann}) this way, 
with the invariant SU(2) measure in the axis-angle parametrization:
\begin{eqnarray}\label{classical1}
\!\!\!\!\!\!\!\!\!\!
&& \Omega_{\Nj}= (2J + 1) \frac{1}{8\pi^2} \int \d\Omega_{\hat{{\bf n}}} 
\int_0^{4\pi} \d \psi \sin^2 \frac{\psi}{2} \,
 D^J_{\lambda\lambda}({\sf R}_{\hat{{\bf n}}}^{-1})  
\left[\prod_{j=1}^{K}\frac{1}{N_j!} \right] \nonumber \\
\!\!\!\!\!\!\!\!\!\!
&& \times \left[ \prod_{\parti=1}^{N}
 \frac{\tr D^{S_{\parti}}({\sf R}_{\hat{{\bf n}}}(\psi))}{(2 \pi \hslash)^3} 
 \int \d^3 {\rm p}_{\parti} 
  \int_V \d^3 {\rm x}_{\parti} \; 
\e^{i {\bf x}_{\parti}/\hslash \cdot ({\bf p}_{\parti}-{\sf R}_{\hat{{\bf n}}}^{-1}
 (\psi){\bf p}_{\parti})} \right] 
 \delta^4 \left(P - \sum_{\parti=1}^{N} p_\parti \right). \nonumber \\
\!\!\!\!\!\!\!\!\!\! && 
\end{eqnarray}
Setting:
$$
{\bf y} \equiv \frac{{\bf x}}{\hslash}
$$
the eq.~(\ref{classical1}) can be rewritten as:
\begin{eqnarray}\label{classical2}
\!\!\!\!\!\!\!\!\!\!
&& \Omega_{\Nj}= (2J + 1) \frac{1}{8\pi^2} \int \d\Omega_{\hat{{\bf n}}} 
 \int_0^{4\pi} \d \psi \; \sin^2 \frac{\psi}{2} \, 
 D^J_{\lambda\lambda}({\sf R}_{\hat{{\bf n}}}(\psi)^{-1})  
\left[\prod_{j=1}^{K}\frac{1}{N_j!} \right] \nonumber \\
\!\!\!\!\!\!\!\!\!\!
&& \times \left[ \prod_{\parti=1}^{N}
  \frac{\tr D^{S_{\parti}}({\sf R}_{\hat{{\bf n}}}(\psi))}{(2 \pi)^3} 
  \int \d^3 {\rm p}_{\parti} \int_{V/\hslash^3} \d^3 {\rm y}_{\parti} \; 
\e^{i {\bf y}_{\parti} \cdot ({\bf p}_{\parti}-{\sf R}_{\hat{{\bf n}}}^{-1}
 (\psi){\bf p}_{\parti})} \right] 
 \delta^4 \left(P - \sum_{\parti=1}^{N} p_\parti \right). \nonumber \\
\!\!\!\!\!\!\!\!\!\! && 
\end{eqnarray}
In the limit $\hslash \to 0$, the integration domain $V/\hslash^3$ of the Fourier 
integrals in ${\bf y}$ becomes very large and the integrals tend to a Dirac's 
delta distribution. Hence 
$({\bf p}_{\parti} \to {\sf R}_{\hat{{\bf n}}}^{-1} (\psi){\bf p}_{\parti})$
for each ${\bf p}_{\parti}$ and this means that the rotation 
${\sf R}_{\hat{{\bf n}}}^{-1}$ tends to the identity, or, equivalently 
$\psi \to 0, 2\pi, 4\pi$. Indeed, it can be shown that for increasingly small
$\hslash$, the integrand in eq.~(\ref{classical2}) develops 4 symmetric narrow
gaussian peaks in $\psi$ with maxima at $\psi = \epsilon, 2\pi-\epsilon,
2\pi+\epsilon, 4\pi-\epsilon$ with $\epsilon \to 0$ as $\hslash \to 0$.
Hence, we can reduce the integration on the angle $\psi$ to the interval $[0,\pi]$ 
multiplying (\ref{classical2}) by 4 and in this interval approximate $\sin \psi
\simeq \psi$ and $\cos \psi \simeq 1$. If ${\bf v}$ is a generic vector, the 
vector ${\sf R}_{\hat{{\bf n}}}(\psi){\bf v}$ reads:
\begin{equation}\label{rotazvett}
{\sf R}_{\hat{{\bf n}}}(\psi){\bf v}={\bf v} \cos \psi  + (\hat{{\bf n}} 
\times {\bf v}) \sin \psi +(1-\cos \psi ){\bf v}\cdot\hat{{\bf n}}\,\hat{{\bf n}}
\end{equation}
therefore, for small $\psi$:
\begin{equation}\label{rotazvett2}
{\sf R}_{\hat{{\bf n}}}(\psi) \simeq {\bf v}  +  \psi (\hat{{\bf n}} 
\times {\bf v})
\end{equation}
thus:
\begin{equation}\label{rotapprox}
 {\bf p}_{\parti} - {\sf R}_{\hat{{\bf n}}}^{-1} (\psi){\bf p}_{\parti} \simeq
   \psi (\hat{{\bf n}} \times {\bf p})
\end{equation}
By using (\ref{rotapprox}) and $\sin \psi \simeq \psi$ in (\ref{classical2}) we 
obtain:
\begin{eqnarray}\label{classical3}
\!\!\!\!\!\!\!\!\!\!
&& \Omega_{\Nj} \underset{\hslash \to 0}{\simeq} 
 (2J + 1) \frac{4}{8\pi^2} \int \d\Omega_{\hat{{\bf n}}} \int_0^{\pi} \d \psi
 \; \frac{\psi^2}{4} \, D^J_{\lambda  \lambda}({\sf R}_{\hat{{\bf n}}}^{-1})  
\left[\prod_{j=1}^{K}\frac{1}{N_j!} \right] \nonumber \\
\!\!\!\!\!\!\!\!\!\!
&& \times \left[ \prod_{\parti=1}^{N}\frac{1}{(2 \pi)^3} \, 
 \tr D^{S_{\parti}}({\sf R}_{\hat{{\bf n}}}(\psi))\int \d^3 {\rm p}_{\parti} 
  \int_{V/\hslash^3} \d^3 {\rm y}_{\parti} \; 
\e^{i {\bf y}_{\parti} \cdot \psi(\hat{{\bf n}} \times {\bf p})} \right] 
 \delta^4 \left(P - \sum_{\parti=1}^{N} p_\parti \right) \; . \nonumber \\
\!\!\!\!\!\!\!\!\!\! &&
\end{eqnarray}
Now $\tr D^{S_{\parti}}({\sf R}_{\hat{{\bf n}}})$ can be replaced with 
$(2S_\parti + 1)$ if $\psi \to 0$. On the other hand, if $J/\hslash$ and $\lambda/\hslash$ 
are macroscopic integer numbers, we can also take the large spin limit of the
rotation matrix and replace $D^J_{\lambda \lambda}({\sf R}_{\hat{{\bf n}}}(\psi)^{-1})$ 
with $\exp[i \psi \hat{\bf n} \cdot {\bf J}/\hslash]$ where
 ${\bf J} = \lambda \hat{\bf k}$ is defined as the total angular momentum vector. 
Therefore: 
\begin{eqnarray}\label{classical4}
&& \Omega_{\Nj} \underset{\hslash \to 0}{\simeq} 
 (2J + 1) \frac{1}{8\pi^2} \int \d\Omega_{\hat{{\bf n}}} \int_0^{\pi} 
 \d \psi \; \psi^2 \; \e^{i \psi \hat{\bf n} \cdot {\bf J}/\hslash} 
 \left[\prod_{j=1}^{K}\frac{2S_j+1}{N_j!} \right] \nonumber \\
&& \times \left[ \prod_{\parti=1}^{N}\frac{1}{(2 \pi)^3} 
  \int \d^3 {\rm p}_{\parti} \int_{V/\hslash^3} \d^3 {\rm y}_{\parti} \; 
\e^{i {\bf y}_{\parti} \cdot \psi(\hat{{\bf n}} \times {\bf p})} \right] 
 \delta^4 \left(P - \sum_{\parti=1}^{N} p_\parti \right) \; .
\end{eqnarray}
We can introduce a classical orbital momentum ${\bf L}$ by noting that: 
\begin{equation}\label{eqn:genclass4}
{\bf y}_{\parti_j} \cdot (\hat{{\bf n}} \times {\bf p}_{\parti_j})=-\hat{{\bf n}} 
\cdot ({\bf y}_{\parti_j} \times {\bf p}_{\parti_j}) = - 
\hat{{\bf n}} \cdot {\bf L}_{\parti_j}/ \hslash
\end{equation}
Also, one can introduce a the three-vector $\phiv = \psi \hat{{\bf n}}$ and 
rewriting the eq.~(\ref{classical4}) with $\d^3 \phi \equiv \d 
\Omega_{\hat{{\bf n}}}\d \psi \psi^2$:
\begin{eqnarray}\label{classical5}
&&\Omega_{\Nj} \simeq (2J + 1) \frac{1}{8 \pi^2} \int \d^3 \phi \; 
\e^{i \phiv \cdot{\bf J}/\hslash} \left[\prod_{j=1}^{K}\frac{(2S_j+1)^{N_j}}{N_j!} \right]
\left[ \prod_{\parti=1}^{N} \int \d^3 {\rm p}_{\parti} \right. \; \nonumber \\
&& \left. \times \frac{1}{(2 \pi)^3} \int_{V/\hslash^3} \d^3 {\rm y}_{\parti} \; 
 \e^{-i \phiv \cdot {\bf L}_{\parti}/\hslash} \right]
 \delta^4 \left(P - \sum_{\parti=1}^{N} p_\parti \right)
\end{eqnarray}
where the integration in $\phiv$ is performed over the whole space as the contribution
of the $\psi$ interval $[\pi,+\infty)$ is negligible for $\hslash \to 0$. Now, 
by restoring the variable ${\bf x}=\hslash {\bf y}$ and rescaling $\phiv \to \phiv/\hslash$
we finally obtain:
\begin{eqnarray}\label{classical6}
&& \Omega^{\Nj}_{\rm classical} = (2J + 1) \frac{\pi}{\hslash^{3N-3}} 
\left[ \prod_{j=1}^{K}\frac{(2S_{j} + 1)^{N_j}}{N_j!}\right]  
 \nonumber \\
&\times& \left[ \prod_{\parti=1}^{N} \frac{1}{(2\pi)^3} \int \d^3 {\rm p}_{\parti} 
 \int_V \d^3 {\rm x}_{\parti}\right] \; \delta^4 \left(P - \sum_{\parti=1}^{N} 
p_\parti \right)\delta^3\left({\bf J} - \sum_{\parti=1}^{N} {\bf L}_\parti \right) \; .
\end{eqnarray}
which is the classical expression of the microcanonical channel weight with
angular momentum conservation \cite{ericson}. As expected, in the classical case, 
this conservation
implies a simple $\delta^3$ as though the angular momentum components were commuting
observables. To be noted that the contribution of spin reduces to an overall 
degeneracy factor.

%******************************************************************************
\section{Grand-canonical limit}
%******************************************************************************

It is also interesting to derive the (grand-)canonical partition function with angular 
momentum conservation as a limiting case of the microcanonical partition function for
large volumes and energies. Again, we will confine ourselves to the case of Boltzmann 
statistics. One can start from the expression (\ref{Boltzmann}) of the microcanonical 
channel weight and disregard the factor $\bra{0}\Pro_V\ket{0}$ as $\Pro_V \to {\sf I}$: 
\begin{eqnarray}\label{canonical1}
\!\!\!\!\!\!\!\!\!\!
&& \Omega_{\Nj} = (2J + 1) \frac{1}{8\pi^2} \int \d\Omega_{\hat{{\bf n}}} 
  \int_0^{4\pi} \d \psi \; \sin^2 \frac{\psi}{2} \, 
  D^J_{\lambda\lambda}({\sf R}_{\hat{\bf n}}(\psi)^{-1})  
 \left[ \prod_{j=1}^K \frac{1}{N_j!} \right]  \nonumber \\ 
\!\!\!\!\!\!\!\!\!\! 
&\times&\left[ \prod_{\parti=1}^{N} \int \d^3 {\rm p}_{n}  \frac{1}{(2\pi)^3} 
 \int \d^3 {\rm x} 
 \e^{i {\bf x}\cdot({\bf p}_{\parti}-{\sf R}_{\hat{\bf n}}(\psi)^{-1}{\bf p}_{\parti})} \, 
 \tr D^{S_n}({\sf R}_{\hat{\bf n}}(\psi)) \right]
 \delta^4 \left(P - \sum_{\parti=1}^{N} p_\parti \right) \nonumber \\
\!\!\!\!\!\!\!\!\!\!  &&
\end{eqnarray}
where we have used the axis-angle parametrization.
We have seen in the previous section, dealing with the classical limit, that for 
large volumes, the integrand in $\psi$ in the above equation develops 4 symmetric 
gaussian peaks in $\psi$ with maxima at $\psi = \epsilon, 2\pi-\epsilon,
2\pi+\epsilon, 4\pi-\epsilon$ with $\epsilon \to 0$ as $V \to \infty$ 
and that the Fourier integrals be approximated by:
\begin{equation}\label{fintasy}
  F_V({\bf p}_\parti - {\sf R}_{\hat{\bf n}}(\psi)^{-1}{\bf p}_\parti) 
  \underset{\psi \to 0}{\simeq} \frac{1}{(2\pi)^3} \int_V \d^3{\rm x}\; \e^{i {\bf x} 
  \cdot \psi(\hat{\bf n} \times {\bf p}_{\parti})} 
\end{equation} 
In view of eqs.~(\ref{rotazvett}),(\ref{rotazvett2}). Thus, plugging (\ref{fintasy}),
into eq.~(\ref{canonical1}) and using the same approximations on the group
integration as in Sect.~8: 
\begin{eqnarray}\label{canonical2}
\!\!\!\!\!\!\!\!\!\!
&& \Omega_{\Nj} = (2J + 1) \frac{1}{8\pi^2} \int \d\Omega_{\hat{{\bf n}}} 
  \int_0^{\pi} \d \psi \; \psi^2 \, 
  D^J_{\lambda\lambda}({\sf R}_{\hat{\bf n}}(\psi)^{-1})  
 \left[ \prod_{j=1}^K \frac{1}{N_j!} \right] \nonumber \\ 
\!\!\!\!\!\!\!\!\!\! 
&\times& \left[ \prod_{\parti=1}^{N} \int_V \d^3 {\rm p}_{\parti}
  \frac{1}{(2\pi)^3} \int_V \d^3 {\rm x}_\parti 
 \e^{-i \psi \hat{\bf n}\cdot({\bf x}_\parti \times{\bf p}_{\parti})} \, 
 \tr D^{S_n}({\sf R}_{\hat{\bf n}}(\psi)) \right]
 \delta^4 \left(P - \sum_{\parti=1}^{N} p_\parti \right) \nonumber \\
\!\!\!\!\!\!\!\!\!\!  &&
\end{eqnarray}

A quick way of deriving the grand-canonical limit of the expression (\ref{canonical2}) 
is to replace the $\delta$ factor of four momentum conservation with the canonical 
weight $\prod_{\parti=1}^N \e^{-\varepsilon_{\parti}/T} = \prod_{j=1}^{K} 
\prod_{\parti_j=1}^{N_j} \e^{-\varepsilon_{\parti_j}/T}$ where $T$ is the temperature. 
In fact, this procedure corresponds to replace the projector $\delta^4(P-{\hat P})$,
with the operator $\exp(-{\hat H}/T)$ so as to match the definition of grand-canonical 
partition function (with vanishing chemical potentials) with fixed angular momentum:
\begin{equation}\label{grandca}
 Z_J = \tr [ \exp(-{\hat H}/T) P_{J\lambda} \Pro_V ]
\end{equation}
Indeed, it can be seen from eqs.~(\ref{micropf2}),(\ref{proji1}) that (\ref{grandca}) 
is obtained from the microcanonical partition function definition replacing the
projector onto fixed energy-momentum with the familiar $\exp[-{\hat H}/T]$. We 
could have also developed an integral expression of the microcanonical partition
function summing up all channels (\ref{canonical2}) and inserting a Fourier expansion
of the $\delta^4$ in order to calculate the grand-canonical limit through a saddle-point
expansion \cite{bf1}, but this would imply lengthy calculations. 
So, with the above replacement, we obtain a grand-canonical channel weight:
\begin{eqnarray}\label{gcanlim1}
\!\!\!\!\!\!\!\!\!\!
&& Z_{\Nj}= (2J + 1) \frac{1}{8\pi^2} \int \d\Omega_{\hat{{\bf n}}} 
  \int_0^{\pi} \d \psi \; \psi^2 \,
  D^J_{\lambda\lambda}({\sf R}_{\hat{\bf n}}(\psi)^{-1}) \nonumber \\
\!\!\!\!\!\!\!\!\!\! 
&&\times \left[\prod_{j=1}^{K}\frac{1}{N_j!} \prod_{\parti_j=1}^{N_j} 
 \int \d^3 {\rm p}_{\parti_j} \frac{1}{(2\pi)^3} \int_V \d^3 {\rm x}_{\parti_j} 
 \e^{-i \psi \hat{\bf n}\cdot({\bf x}_{\parti_j}\times{\bf p}_{\parti_j})} \, 
 \tr D^{S_j}({\sf R}_{\hat{\bf n}}(\psi)) \, \e^{-\varepsilon_{\parti_j}/T} \right] 
\nonumber \\
\!\!\!\!\!\!\!\!\!\!
&& = (2J + 1) \frac{1}{8\pi^2} \int \d\Omega_{\hat{{\bf n}}} 
 \int_0^{\pi} \d \psi \; \psi^2 \,D^J_{\lambda\lambda}({\sf R}_{\hat{\bf n}}(\psi)^{-1})
  \nonumber \\ 
\!\!\!\!\!\!\!\!\!\!
&&\times \left[ \prod_{j=1}^{K} \frac{1}{(2\pi)^3}  \int_V \d^3 {\rm x} 
  \int \d^3 {\rm p} \; \e^{-\varepsilon_j/T} \e^{-i \psi \hat{\bf n}\cdot({\bf x} 
 \times {\bf p})} \, \tr D^{S_j}({\sf R}_{\hat{\bf n}}(\psi)) 
 \right]^{N_j} 
 \; .
\end{eqnarray}
It is now straightforward to calculate the full grand-canonical partition 
function by summing over all possible channels, i.e. over all multiplicities $\Nj$ 
from $0$ to $\infty$. The grand-canonical partition function $Z_J$ with fixed angular 
momentum turns out to be an SU(2) integral of an exponential function:
\begin{eqnarray}\label{gcpfj}
 Z_J&=& \sum_{N_1=0}^\infty \ldots \sum_{N_K=0}^\infty Z_{\Nj} \nonumber \\
    &=& (2J + 1) \frac{1}{8\pi^2} \int \d\Omega_{\hat{{\bf n}}} 
  \int_0^{\pi} \d \psi \; \psi^2 \,
  D^J_{\lambda\lambda}({\sf R}_{\hat{\bf n}}(\psi)^{-1}) \nonumber \\ 
&\times& \exp \left[ \sum_{j=1}^{K} \frac{1}{(2\pi)^3} 
 \int_V \d^3 {\rm x} \int \d^3 {\rm p} \; \e^{-\varepsilon_j/T} 
 \e^{-i \psi \hat{\bf n}\cdot({\bf x} \times {\bf p})} \, 
  \tr D^{S_j}({\sf R}_{\hat{\bf n}}(\psi)) 
 \right] 
\end{eqnarray}
Defining the vector $\phiv = \psi \hat{\bf n}$ (so that $\hat \phiv \equiv \hat{\bf n}$, 
the eq.~(\ref{gcpfj}) can be rewritten as:
\begin{eqnarray}\label{gcpfj2}
 Z_J&=& (2J + 1) \frac{1}{8\pi^2} \int_{|\phiv|<\pi} \d^3 \phiv \; 
  D^J_{\lambda\lambda}({\sf R}_{\hat{\phiv}}(\phi)^{-1}) \nonumber \\ 
 &\times& \exp \left[ \sum_{j=1}^{K} \frac{1}{(2\pi)^3} 
 \int_V \d^3 {\rm x} \int \d^3 {\rm p} \; \e^{-\varepsilon_j/T} 
 \e^{-i \phiv \cdot({\bf x} \times {\bf p})} \, \tr D^{S_j}({\sf R}_{\hat{\phiv}}(\phi)) 
 \right] 
\end{eqnarray}
For large values of $J$ and $\lambda$, we can replace the matrix element in
(\ref{gcpfj2}) with $\exp[i\phiv \cdot {\bf J}]$ where ${\bf J}= \lambda \hat{\bf k}$,
like in Sect.~8 and write $Z_J$ as:
\begin{eqnarray}\label{gcpfj3}
 Z_J&=& (2J + 1) \frac{1}{8\pi^2} \int_{|\phiv|<\pi} \d^3 \phi \; 
  \e^{i \phiv\cdot {\bf J}} \nonumber \\ 
 &\times& \exp \left[ \sum_{j=1}^{K} \frac{1}{(2\pi)^3} 
 \int_V \d^3 {\rm x} \int \d^3 {\rm p} \; \e^{-\varepsilon_j/T} 
 \e^{-i \phiv \cdot({\bf x} \times {\bf p})} \, 
 \tr D^{S_j}({\sf R}_{\hat{\phiv}}(\phi)) 
 \right] 
\end{eqnarray}
which corresponds to a classical limit were not for the presence of the traces
of matrices $D^{S_j}({\sf R}_{\hat{\phiv}}(\phi))$. 

%************************************************************************
\section{Microcanonical ensemble with fixed parity}
%************************************************************************

In this section we will show how to enforce a fixed parity for an ideal 
relativistic quantum gas. 

As has been shown in Sect.~2, the projector $\Pro_i$ onto irreducible Poincar\'e 
states can be factorized as in eq.~(\ref{final}):
\begin{equation}\label{piparity}
 \Pro_i=\Pro_{P J \lambda} \frac{{\sf I} + \Pi \hat{{\sf \Pi}}}{2} 
\end{equation}
where $\Pi$ is the parity of the system. Rewriting the general state weight with the 
full projector:
\begin{equation}\label{piparity2}
\omega_f \equiv \bra{f} \Pro_{P J \lambda} \frac{{\sf I} + 
\Pi \hat{{\sf \Pi}}}{2} \Pro_V \ket{f}= \frac{1}{2}  \bra{f} \Pro_{P J \lambda} \Pro_V 
\ket{f}+ \frac{\Pi}{2}  \bra{f} \Pro_{P J \lambda} \hat{{\sf \Pi}} \Pro_V \ket{f} \qquad
\end{equation}
we are left with the calculation of two terms: the first is simply eq.~(\ref{omegan4}) 
times a factor $1/2$, while the second is a new one to be calculated. To do that,
one should find the action of ${\sf \Pi}$ on a single particle state. 
Indeed, there is no unique answer to this question, as any definition of the unitary
representation of space inversion on Hilbert space fulfilling commutation rules of 
\poinc\ is equally valid. This leaves some freedom and we choose:
\begin{equation}\label{paritydef}
 \hat{{\sf \Pi}} \ket{p, \sigma}=\sum_\mu  \ket{{\sf \Pi}p, \mu} D^{S}_{\mu \sigma}
 ([{\sf \Pi}p]^{-1}{\sf \Pi}[p]) \; .
\end{equation}
where ${\sf \Pi}p=(\varepsilon, -{\bf p})$. To be a good one, the definition 
(\ref{paritydef}) must meet the following requirements:
\begin{itemize}
\item{} $D^{S}([{\sf \Pi}p]^{-1}{\sf \Pi}[p])$ must be unitary;
\item{} for each state $\ket{p^0, \sigma}$, where $p^0=(m,{\bf 0})$, the space 
inversion $\hat{{\sf \Pi}}$ must yield $\hat{{\sf \Pi}} \ket{p^0, \sigma}=  
\ket{p^0, \sigma}\eta$ where $\eta$ is the intrinsic parity.
\end{itemize}
With regard to the first point, it should be noted that $D^S$ is a finite dimensional
representation of the universal covering group of SO(1,3)$_+^\uparrow$, i.e. SL(2,C)
and a unitary representation of the subgroups SO(3) and SU(2) respectively. For it
to be extended so as to include space inversion operator, one should recall the second
Shur's lemma:
\begin{center}
{\em In a finite-dimensional irreducible representation of a group G, the only 
elements which commute with all others are multiples of the identity}. 
\end{center}
Since space inversion commutes with all rotations, i.e. $[{\sf \Pi},{\sf R}]=0$, 
then:
\begin{displaymath}
 D^{S}({\sf \Pi})D^{S}({\sf R})=D^{S}({\sf \Pi}{\sf R})=D^{S}({\sf R}{\sf \Pi})=
 D^{S}({\sf R})D^{S}({\sf \Pi})
\end{displaymath}
thus $D^S({\sf \Pi})$ commutes with all elements of a finite-dimensional 
representation of a group and, according to the second Schur's lemma, must be a 
multiple of the identity:
\begin{equation}\label{parityrep}
  D^{S}({\sf \Pi})= \eta \, {\rm  I}
\end{equation}
Moreover, since $D^{S}({\sf \Pi}^2)=D^{S}( \, {\sf I}\, )= \eta^2 \, I$, 
then $\eta^2=1 \Rightarrow \eta= \pm 1$. In the second step of the proof, we
show that $[{\sf \Pi}p]^{-1}{\sf \Pi}[p]$ does not involve any boost, hence
it belongs to the subgroup O(3). This can be done by noting that, when acting upon 
the unit time vector $t=(1,{\bf 0})$:
\begin{equation}\label{action}
 [{\sf \Pi}p]^{-1}{\sf \Pi}[p] t \longrightarrow t \qquad 
 \Rightarrow [{\sf \Pi}p]^{-1}{\sf \Pi}[p] \in {\rm O(3)} \; .
\end{equation}
Since the above transformation belongs to O(3), it is either a rotation or the
product of a rotation by ${\sf \Pi}$ itself. In both cases, because of 
(\ref{parityrep}) and $\eta^2 = 1$, the matrix 
$D^S([{\sf \Pi}p]^{-1}{\sf \Pi}[p])$ is unitary and the first requirement is met.

As far as the second requirement is concerned, noting that $[p^0] =[{\sf \Pi}p^0] 
= {\sf I}$ and using (\ref{paritydef}),(\ref{parityrep}):
\begin{eqnarray}
&& \hat{{\sf \Pi}} \ket{p^0, \sigma} = \sum_\mu \ket{{\sf \Pi}p^0, \mu} 
 D^{S}_{\mu \sigma}([{\sf \Pi}p^0]^{-1}{\sf \Pi}[p^0]) = \sum_\mu 
 \ket{p^0, \mu} D^{S}_{\mu \sigma}({\sf \Pi}) \nonumber \\
&=& \sum_\mu \ket{p^0, \mu} \eta \delta_{\mu \sigma}=  \ket{p^0, \sigma} \eta 
\end{eqnarray}
where $\eta$ is the intrinsic parity. Hence, the second requirement is also met.

We now have to check the consistency of (\ref{paritydef}) with commutation rules:
\begin{eqnarray} \label{comrules}
 &&\hat{{\sf \Pi}} \hat{{\sf R}} \ket{p, \sigma}= \hat{{\sf R}} \hat{{\sf \Pi}} 
 \ket{p, \sigma}  \nonumber \\
&&\hat{{\sf \Pi}} \hat{{\sf L}} \ket{p, \sigma}= \hat{{\sf L}}^{-1} \hat{{\sf \Pi}} 
 \ket{p, \sigma} \qquad \qquad\forall \ket{p, \sigma}\; . \\
&&\hat{{\sf \Pi}} \hat{{\bf P}} \ket{p, \sigma}= -\hat{{\bf P}}\hat{{\sf \Pi}} 
  \ket{p, \sigma} \nonumber
\end{eqnarray}
${\sf R}$ being a rotation, ${\sf L}$ a pure Lorentz boost and $\hat {\bf P}$ being
the momentum operator. The last equality is trivially fulfilled. For the first 
commutation rule, going back to definition~(\ref{paritydef}):
\begin{eqnarray} \label{wigner1}
 &&\hat{{\sf \Pi}} \hat{{\sf R}} \ket{p, \sigma}= \sum_\tau
 \hat{{\sf \Pi}}\ket{{\sf R} p, \tau} D^{S}_{\tau  \sigma}([{\sf R}p]^{-1}{\sf R}[p] )
 \nonumber \\
&=& \sum_{\mu,\tau} \ket{{\sf \Pi}{\sf R}p, \mu} D^{S}_{\mu \tau } ([{\sf \Pi}{\sf R}p]^{-1} 
 {\sf \Pi}[{\sf R} p]) D^{S}_{\tau  \sigma}([{\sf R} p]^{-1}{\sf R}[p] ) \nonumber \\
&=& \sum_{\mu}
 \ket{{\sf \Pi}{\sf R}p, \mu} D^{S}_{\mu \sigma } ([{\sf \Pi}{\sf R}p]^{-1} {\sf \Pi}{\sf R}[p] )
\end{eqnarray}
This relation tells us that the composition of operators transforming a state with 
a Wigner-rotation's rule (like space inversion itself according to 
(\ref{paritydef})) yields an operator fulfilling the same rule. Hence we have:
\begin{equation} \label{wigner2}
\hat{{\sf R}}\hat{{\sf \Pi}}  \ket{p, \sigma} = \sum_\mu \ket{{\sf R}{\sf \Pi}p, \mu} 
D^{S}_{\mu \sigma } ([{\sf R}{\sf \Pi}p]^{-1}{\sf R} {\sf \Pi}[p] )
\end{equation}
and taking into account that $[{\sf \Pi},{\sf R}]=0$ it follows from eqs.~(\ref{wigner1})
and (\ref{wigner2}) that $\hat{{\sf \Pi}} \hat{{\sf R}} \ket{p, \sigma} = 
\hat{{\sf R}}\hat{{\sf \Pi}} \ket{p, \sigma}$ $\forall \ket{p, \sigma}$. 

Similarly, for Lorentz boosts:
\begin{eqnarray} \label{boost}
&&\hat{{\sf \Pi}}\hat{{\sf L}}  \ket{p, \sigma} = \sum_\mu \ket{{\sf \Pi}{\sf L} p, \mu}
D^{S}_{\mu \sigma } ([{\sf \Pi}{\sf L} p]^{-1}{\sf \Pi}{\sf L}[p] ) \nonumber \\
&&\hat{{\sf L}}^{-1}\hat{{\sf \Pi}}  \ket{p, \sigma} = \sum_\mu \ket{{\sf L}^{-1}{\sf \Pi}p, \mu}
D^{S}_{\mu \sigma } ([{\sf L}^{-1}{\sf \Pi}p]^{-1}{\sf L}^{-1}{\sf \Pi}[p] )
\end{eqnarray}
thus, taking into account that  
${\sf \Pi}{\sf L} p={\sf L}^{-1}{\sf \Pi} p$ also the last equality in~(\ref{comrules}) 
is proved.

The conclusion is that the (\ref{paritydef}) is consistent with all properties 
needed by a good definition of space inversion representation. 

We can now turn to calculate the second term in eq.~(\ref{piparity2}). We will 
consider the special case of a spherical system, so that $[\hat{{\sf \Pi}},\Pro_V]=0$, 
and considering the simple case of a single particle state:
\begin{equation} \label{omegapar}
\bra{p, \sigma } \Pro_{P J \lambda} \hat{{\sf \Pi}} \Pro_V \ket{p, \sigma}
\end{equation}
Since $[\hat{{\sf \Pi}}, \Pro_V]=0$, we can rewrite (\ref{omegapar}) as:
\begin{eqnarray} \label{omegapar2}
\sum_{\tau} \bra{p, \sigma } \Pro_{P J \lambda} \Pro_V \ket{{\sf \Pi}p, \tau} 
D^{S}_{\tau  \sigma }([{\sf \Pi}p]^{-1}{\sf \Pi}[p])
\end{eqnarray}
Inserting a resolution of the identity:
\begin{eqnarray} \label{omegapar3}
\sum_{\tau,\nu} \int \d^3 {\rm p}' \bra{p, \sigma } \Pro_{P J \lambda} 
 \ket{p', \nu} \bra{p', \nu} \Pro_V \ket{{\sf \Pi}p, \tau}
 D^{S}_{\tau  \sigma }([{\sf \Pi}p]^{-1}{\sf \Pi}[p]) \; .
\end{eqnarray}
and, by using (\ref{matrixsingle}) and (\ref{expand2}) we can write:
\begin{eqnarray}\label{omegapar4}
\!\!\!\!\!\!\!\!\!\!\!
 &&\bra{p,\sigma} \Pro_{P J \lambda} \hat{{\sf \Pi}} \Pro_V \ket{p, \sigma}=
 \delta^4(P - p) \int \d^3 {\rm p}' \; (2J + 1) \int \d{\sf R} \; 
 D^{J}_{\lambda\lambda}({\sf R}^{-1})\delta^3({\sf R}{\bf p}'-{\bf p})\nonumber \\   
\!\!\!\!\!\!\!\!\!\!\!&&\times 
 \sum_{\nu,\tau} D^S_{\sigma\nu}([{\sf R}p']^{-1}{\sf R}[p'])\frac{1}{2} 
 \left( D^S_{\nu\tau}([p']^{-1}[{\sf \Pi}p])+ D^S_{\nu\tau}([p']^\dagger
 [{\sf \Pi}p]^{\dagger-1})\right)
 D^{S}_{\tau\sigma}([{\sf \Pi}p]^{-1}{\sf \Pi}[p]) \nonumber \\
\!\!\!\!\!\!\!\!\!\!\!&& \times F_V({\bf p}+{\bf p'}) \bra{0}\Pro_V\ket{0}
\end{eqnarray}
The matrix products in (\ref{omegapar4}) can be worked out as follows:
\begin{eqnarray}\label{matprod}
&& \sum_{\nu,\tau} D^S_{\sigma\nu}([{\sf R}p']^{-1}{\sf R}[p']) 
 D^S_{\nu\tau}([p']^{-1}[{\sf \Pi}p]) D^{S}_{\tau\sigma}([{\sf \Pi}p]^{-1}{\sf \Pi}[p]) 
 = D^S_{\sigma\sigma}([p]^{-1}{\sf R \Pi}[p]) \nonumber \\
&& \sum_{\nu,\tau} D^S_{\sigma\nu}([{\sf R}p']^{-1}{\sf R}[p'])
  D^S_{\nu\tau}([p']^\dagger[{\sf \Pi}p]^{\dagger-1})
 D^{S}_{\tau\sigma}([{\sf \Pi}p]^{-1}{\sf \Pi}[p]) \nonumber \\
&=& \sum_{\nu,\tau} D^S_{\sigma\nu}([{\sf R}p']^{\dagger}{\sf R}^{\dagger-1}
  [p']^{\dagger-1}) D^S_{\nu\tau}([p']^\dagger [{\sf \Pi}p]^{\dagger-1})
 D^{S}_{\tau\sigma}([{\sf \Pi}p]^{-1}{\sf \Pi}[p])^{\dagger-1} 
\nonumber \\
&=& \sum_{\tau} D^S_{\sigma\tau}([{\sf R}p']^{\dagger}{\sf R}
  [{\sf \Pi}p]^{\dagger-1})
  \left( D^{S}([{\sf \Pi}p]^{\dagger}) D^S({\sf \Pi})^{\dagger-1} 
  D^S([p]^{\dagger-1}) \right)_{\tau\sigma} \nonumber \\
&=& \sum_{\tau} D^S_{\sigma\tau}([{\sf R}p']^{\dagger}{\sf R})
  \left(D^S({\sf \Pi}) D^S([p]^{\dagger-1})\right)_{\tau\sigma} \nonumber \\
&=& D^S_{\sigma\sigma}([p]^{\dagger}{\sf R \Pi}[p]^{\dagger-1})
\end{eqnarray}
where the constraint ${\sf R}p'=p$ has been used as well as the unitarity of the
Wigner rotation and of the matrices $D^{S}([{\sf \Pi}p]^{-1}{\sf \Pi}[p])$
and $D^S({\sf \Pi})$.
Therefore, taking (\ref{matprod}) into account, the eq.~(\ref{omegapar4}) 
can be simplified into:
\begin{eqnarray}\label{omegapar5}
\!\!\!\!\!\!\!\!\!\!\!
&&\bra{p,\sigma} \Pro_{P J \lambda} \hat{{\sf \Pi}} \Pro_V \ket{p, \sigma}=
 \delta^4(P - p) \int \d^3 {\rm p}' \; (2J + 1) \int \d{\sf R} \; 
 D^{J}_{\lambda\lambda}({\sf R}^{-1}) \delta^3({\sf R}{\bf p}'-{\bf p})\nonumber \\   
\!\!\!\!\!\!\!\!\!\!\!&&\times 
  \frac{1}{2} \left( D^S([p]^{-1}{\sf R \Pi}[p])
+ D^S([p]^{\dagger}{\sf R \Pi}[p]^{\dagger-1} \right)_{\sigma\sigma} 
 F_V({\bf p}+{\bf p'}) \bra{0}\Pro_V\ket{0}
\end{eqnarray}
If we now sum over polarization states $\sigma$, the last factor in previous 
equation yields the trace of $D^S({\sf R \Pi})$.
Since:
\begin{equation} \label{tracepar}
 \tr D^S({\sf R}{\sf \Pi}) = \tr \left[ D^S({\sf R}) D^S({\sf \Pi}) \right]
 =\tr \left[ D^S({\sf R})\eta \, I \right] = \eta \, \tr D^S({\sf R}) \; . 
\end{equation}
we finally get:
\begin{eqnarray} \label{omegapar6}
 \sum_{\sigma} \bra{p, \sigma } \Pro_{P J \lambda} \hat{{\sf \Pi}} \Pro_V 
 \ket{p,\sigma} &=& 
 \eta \, \delta^4( P - p) (2J + 1) \int \d{\sf R} \; 
 D^{J}_{\lambda\lambda}({\sf R}^{-1})  \nonumber \\   
&\times& F_V({\bf p}+{\sf R}^{-1}{\bf p}) \tr D^S({\sf R}) \bra{0}\Pro_V\ket{0}\; . 
\end{eqnarray}
This is, up to a factor $\eta $, exactly the same result one would obtain without 
space inversion, the only difference being in the argument Fourier integral which 
is now $({\bf p}+{\sf R}^{-1}{\bf p})$ instead of $({\bf p}-{\sf R}^{-1}{\bf p})$. 
Thus, it is not difficult to realize that the contribution to the microcanonical 
weight of a generic channel with $N$ different particles, of the space-reflected 
term in the right-hand side of (\ref{piparity2}) coincides with eq.~(\ref{chanweight}) 
up to a factor $\Pi \prod_{j=1}^{k} \eta_j^{N_j} \equiv \Pi\Pi_{\Nj}$ and a ``plus'' 
sign in the argument of Fourier integrals. Finally, we quote the full expression 
of the microcanonical channel weight for a spherical system at rest in the 
Boltzmann statistics:
\begin{eqnarray}\label{paritychannel}
 \Omega_{\Nj} &=& \frac{1}{4\pi} \int_0^{4\pi} \d \psi \; \sin {\psi \over 2}
 \sin \left(J+\frac{1}{2}\right) \psi \left[ \prod_{j=1}^K \frac{1}{N_j!} 
 \prod_{\parti_j=1}^{N_j} \right. \int \d^3{\rm p}_{n_j} \nonumber \\  
  &\times& \frac{1}{2} 
 \left( F_V({\bf p}_{\parti_j}- {\sf R}_3(\psi)^{-1}{\bf p}_{\parti_j}) + 
  \Pi \Pi_{\Nj} F_V({\bf p}_{\parti_j} + {\sf R}_3(\psi)^{-1}{\bf p}_{\parti_j}) \right)
 \nonumber \\
 &\times & \left.
 \left[\frac{\sin(S_j+\frac{1}{2})\psi}{\sin(\frac{\psi}{2})}\right]
 \right] \delta^4 \left(P - \sum_{\parti=1}^{N} p_\parti \right) \bra{0} \Pro_V \ket{0}
\end{eqnarray}
which can be obtained by removing all permutations but identity in 
eq.~(\ref{spherical1})
and adding the parity-transformed term. The microcanonical partition function can 
be obtained by summing (\ref{paritychannel}) over all possible channels.  

%*************************************************************************
\section{Conclusions}
%*************************************************************************

We have studied the microcanonical ensemble of a multi-species ideal relativistic 
quantum gas fixing the maximal set of observables pertaining to the space-time symmetry
group. In the rest frame of the system, where ${\bf P}=0$ this means enforcing 
the conservation of energy-momentum, total angular momentum and parity. We have 
provided a full consistent treatment and solved the problem for the most general case
of particles with spin, generalizing the results of Cerulus \cite{cerang}.

The microcanonical partition function $\Omega$ has been expressed as a sum
over channels, i.e. at fixed multiplicities of each particle species, each
term being defined microcanonical channel weight. These have been obtained in a 
quantum field theoretical framework, applying to spinorial fields the formulae 
derived for scalar fields in our previous work \cite{micro1}.
This extension relies on two reasonable, yet unproved assumptions:
\begin{enumerate}
\item{} the validity of eqs.~(\ref{keypart}) and (\ref{keyapart}) which is 
conjectured on the basis of the $V \to \infty$ limit and the fact that they have 
been proved for the scalar field \cite{micro1};
\item{} that also for particles with spin, the expressions of the channel 
weights involving identical particles are the same as those with different
particles with (anti)symmetrization, that has been proved for the scalar field
\cite{micro1}.
\end{enumerate} 

The implementation of angular momentum conservation 
leads to an expression of the microcanonical weight of a channel involving a 
further integration over the parameters of the SU(2) group, with respect to 
the case of only energy-momentum conservation. This expression nicely agrees 
with that obtained with an expansion onto the angular momentum basis and leads
to the expected classical limit, already known and used in literature. We have 
been working out the grand-canonical partition function with fixed angular momentum 
in both the case of small and large angular momentum. We also obtained the 
microcanonical partition function of a system with fixed parity.

%**********************************************************************************

\end{document}